\pdfoutput=1
\documentclass[12pt]{article}
 \usepackage[margin=1in]{geometry}
 \usepackage{bm}
\usepackage{cite}
\usepackage{amsmath}
\usepackage{graphicx,psfrag,epsf}
\usepackage{enumerate}
\usepackage[authoryear]{natbib}
\setcitestyle{authoryear, open={(}, close={)}, citecolor=black}
\usepackage{multibib}
\newcites{supp}{Supplementary References}
\usepackage{url}

    \usepackage{caption}
    \usepackage{subcaption}
    \usepackage{float}
    \usepackage{hyperref}
    \hypersetup{
        colorlinks=true,
        linkcolor=blue,
        filecolor=magenta,      
        urlcolor=cyan,
        citecolor = black
    }
     
    \usepackage{xcolor}
    \usepackage{cancel}
    \usepackage{listings}
    \usepackage{amssymb}
    \usepackage{amsthm}
    \usepackage{breqn}
    \usepackage{enumerate}	
    \usepackage{wrapfig}
    \usepackage{graphicx}
    \graphicspath{ {./images/} }
    \usepackage{titlesec}
    \titlespacing{\section}{0pt}{\parskip}{-\parskip}
\titlespacing{\subsection}{0pt}{\parskip}{-\parskip}
\titlespacing{\subsubsection}{0pt}{\parskip}{-\parskip}
    
\usepackage{tikz}
\usetikzlibrary{shapes}
\usetikzlibrary{fit}
\usetikzlibrary{chains}
\usetikzlibrary{arrows}

\tikzstyle{latent} = [circle,fill=white,draw=black,inner sep=1pt,
minimum size=20pt, font=\fontsize{10}{10}\selectfont, node distance=1]
\tikzstyle{obs} = [latent,fill=gray!25]
\tikzstyle{const} = [rectangle, inner sep=0pt, node distance=1]
\tikzstyle{factor} = [rectangle, fill=black,minimum size=5pt, inner
sep=0pt, node distance=0.4]
\tikzstyle{det} = [latent, diamond]

\tikzstyle{plate} = [draw, rectangle, rounded corners, fit=#1]
\tikzstyle{wrap} = [inner sep=0pt, fit=#1]
\tikzstyle{gate} = [draw, rectangle, dashed, fit=#1]

\tikzstyle{caption} = [font=\footnotesize, node distance=0] %
\tikzstyle{plate caption} = [caption, node distance=0, inner sep=0pt,
below left=5pt and 0pt of #1.south east] %
\tikzstyle{factor caption} = [caption] %
\tikzstyle{every label} += [caption] %

    \theoremstyle{definition}

    \theoremstyle{remark}

\theoremstyle{definition}

\renewcommand{\epsilon}{\varepsilon}

\newcommand{\iid}{\stackrel{\mathrm{iid}}{\sim}}
\newcommand{\ind}{\stackrel{\mathrm{ind}}{\sim}}

\newcommand{\z}{\mathbf{z}}

\newcommand{\w}{\mathbf{w}}
\newcommand{\x}{\mathbf{x}}

\renewcommand{\epsilon}{\varepsilon}

\newcommand{\W}{\Omega} 

\usepackage{listings}
\usepackage{xcolor}
\newcommand{\blind}{1}

\begin{document}

\def\spacingset#1{\renewcommand{\baselinestretch}%
{#1}\small\normalsize} \spacingset{1}

%%%%%%%%%%%%%%%%%%%%%%%%%%%%%%%%%%%%%%%%%%%%%%%%%%%%%%%%%%%%%%%%%%%%%%%%%%%%%%

\if1\blind
{
  \title{\bf Microbiome subcommunity learning with logistic-tree normal latent Dirichlet allocation}
  \author{Patrick LeBlanc\\
    Department of Statistical Sciences, Duke University\\
    and \\
    Li Ma\thanks{Email: li.ma@duke.edu}\\
    Department of Statistical Sciences,Duke University}
  \maketitle
} \fi

\if0\blind
{
  \bigskip
  \bigskip
  \bigskip
  \begin{center}
    {\LARGE\bf Title}
\end{center}
  \medskip
} \fi

\bigskip
\begin{abstract}
Mixed-membership (MM) models such as Latent Dirichlet Allocation (LDA) have been applied to microbiome compositional data to identify latent subcommunities of microbial species. These subcommunities are informative for understanding the biological interplay of microbes and for predicting health outcomes. However, microbiome compositions typically display substantial cross-sample heterogeneities in subcommunity compositions---that is, the variability in the proportions of microbes in shared subcommunities across samples---which is not accounted for in prior analyses.  As a result, LDA can produce inference which is highly sensitive to the specification of the number of subcommunities and often divides a single subcommunity into multiple artificial ones. To address this limitation, we incorporate the logistic-tree normal (LTN) model into LDA to form a new MM model.  This model allows cross-sample variation in the composition of each subcommunity around some ``centroid'' composition that defines the subcommunity. Incorporation of auxiliary P\'olya-Gamma variables enables a computationally efficient collapsed blocked Gibbs sampler to carry out Bayesian inference under this model.  By accounting for such heterogeneity, our new model restores the robustness of the inference in the specification of the number of subcommunities and allows meaningful subcommunities to be identified.
\end{abstract}

\noindent%
\vfill

\newpage
\spacingset{2}

\section{Introduction}

The human gut microbiome is the genetic content of all bacteria, archaea, viruses, and eukaryotic microbes residing in the human gut and is commonly used to profile the composition of the gut microbiota.  Advances in next-generation sequencing techniques have substantially reduced the cost of this approach and made it widely accessible.  One cost-effective microbiome profiling strategy is based on targeting a single marker gene, the 16S ribosomal RNA (rRNA) gene, through amplicon-based sequencing \citep{2015_Li}. A more expensive, but more precise, approach is whole-genome shotgun metagenomic sequencing \citep{1997WeberMyers}.  Traditionally, sequencing reads have been clustered into Operational Taxonomic Units (OTUs), which serve as the basic unit of microbial taxa. Recently, amplicon sequencing variants (ASVs) have come into wider use as they can achieve more precise characterization of microbial species and resolve the sample-specificity issue of the OTU \citep{2017_Callahan}.  Our work is applicable to either method of characterizing microbial taxa; in the following we shall generically refer to the basic unit as ASVs.

Gut microbiome studies often involve highly heterogeneous samples due to the multitude of factors that can influence an individual's gut microbiota. A useful data analytical strategy for microbiome compositions is to sort microbiome samples into clusters characterized by particular compositional signatures.  In the context of gut microbiome, these clusters are called ``enterotypes'' \citep{2011SiezenKleerebezem} and are associated with health outcomes \citep{2014_Chierico}.  One of the most popular microbiome clustering methods is the Dirichlet-multinomial mixture (DMM) model \citep{2000_Nigam,2012_Holmes}, which uses a hierarchical structure to allow within-cluster cross-sample variability in subcommunity compositions.  However, the DMM is too restrictive to realistically characterize the within-cluster cross-sample variance in microbiome data \citep{2017_Wang,2018_Tang} as it uses a single scalar parameter to characterize the entire covariance structure across all microbial taxa.  More general methods have recently been introduced to alleviate, though not eliminate, this limitation through the use of Dirichlet-tree models \citep{1991Dennis,2017_Wang}. 

Such clustering analysis, however, makes the implicit assumption that each microbiome sample must belong to a {\em single} signature ``community'' characterized by the cluster centroid. This assumption is often unrealistic and overly restrictive for complex environments such as the gut microbiome \citep{2012_Holmes, 2020_Mao}. Recent developments embrace the more relaxed biological hypothesis that the ASVs characterizing a microbiota sample hail from a combination of multiple microbial ``clusters'', or more precisely ``subcommunities''. 

Mixed-membership (MM) models are generalizations of clustering models that provide a generative modeling framework for data involving subcommunity structure as they allow each sample to be composed of multiple subcommunities.  \cite{2019_Sankaran} applied the most well-known MM model, latent Dirichlet allocation (LDA), to microbiome profiling.  Earlier, \cite{2015_Shafei} and \cite{2020_Deek} proposed variations of LDA accounting for environmental factors and inflated zero-counts, respectively, in the microbiome context.

The key motivation for our paper is the observation that existing MM models such as LDA and its variations---originally developed for other contexts such as topic modeling \citep{2003_Blei} and population genetics \citep{2000_Pritchard}---do not incorporate key features of microbiome compositions. Most notably, they assume that a microbial subcommunity's composition must remain {\em exactly the same} across all samples. This is unrealistic in the vast majority of microbiome studies collected from diverse environments such as the gut where samples often possess large heterogeneities  \citep{2018_Tang,2021JeganathanHolmes}. It is interesting to note that such heterogeneity has been well-recognized in clustering models for microbiome data \citep{2012_Holmes,2020_Mao}, but has been largely ignored in existing MM models.  Additionally, choosing the number of subcommunities for LDA is not trivial in the presence of cross-sample heterogeneity, and LDA-based approaches often lead to overestimates in the number of subcommunities in microbiome applications \citep{2021Fukuyamaetal}.  

We introduce a generalization of LDA that aims to appropriately incorporate cross-sample heterogeneity, or  ``random effects", in microbiomal subcommunity compositions due to unmeasured sources, thereby leading to more accurate identification of subcommunities in MM models.  Our approach takes advantage of the availability of a natural tree structure relating the microbial taxa---the phylogenetic tree---which allows us to decompose the compositional vector into a collection of binomial observations on the tree nodes.  This transform serves two purposes. First, it allows us to model the heterogeneity by modeling the vector of log-odds transforms of the binomial probabilities at each node as Gaussian. By modeling the subcommunity compositions as realizations from this logistic-tree normal (LTN) \citep{2021_Wang} distribution, we are able to impose constraints on the underlying covariance structure to ensure the identifiability of the subcommunities.  A second purpose of the tree-based transform is computational. By utilizing the P\'olya-Gamma (PG) data augmentation technique \citep{2013_Polson}, Bayesian inference under the resulting MM model can be readily accomplished through fully conjugate collapsed blocked Gibbs sampling. We term our new model logistic-tree normal latent Dirichlet allocation (LTN-LDA). 

Several other relevant prior works are worth mentioning. Graph-Sparse LDA \citep{2015DoshiVelezetal}  also incorporates random effects from subcommunity-to-subcommunity using a tree structure. However, in the context of microbiome compositions, it would assume that every node of the tree is an ASV which can occur in a sample and is thus incompatible with the phylogenetic tree. Other tree-based MM methods include \cite{2007TamSchultz}, which uses trees to model the abundance of subcommunities in samples, and \cite{2009Andrzejewskietal}, which uses mixtures of trees to model subcommunity composition by explicitly modelling which ASVs must co-occur and which cannot. 

In the following, we will briefly review the LDA and LTN models before introducing the LTN-LDA model.  We will augment the LTN-LDA model using a class of auxiliary P\'olya-Gamma variables \citep{2013_Polson} and present a collapsed blocked Gibbs sampler for carrying out fully Bayesian inference.  We will demonstrate in simulations that, in the presence of cross-sample heterogeneity, inference by LTN-LDA is robust with respect to overspecifying the number of subcommunities while inference by LDA can be highly sensitive to the choice of the number of subcommunities.  We apply LTN-LDA to the dataset of \cite{2011_Dethlefsen}, which has been used for demonstrating MM models in the microbiome settings \citep{2019_Sankaran}, and compare our results to LDA.

\section{Methods}

\subsection{Latent Dirichlet allocation}

Let there be $D$ samples consisting of counts of $V$ unique ASVs indexed by $1,2,\ldots,V$. For sample $d$, let $\x_d = (x_{d,1} ,\dots, x_{d,V})$  be the vector of ASV counts such that $x_{d,v}$ is the total count for ASV $v$ in sample $d$.  Let $N_{d} = \sum_{v=1}^V x_{d,v}$ be the sum of counts in sample $d$, which is determined by the sequencing depth.  Subcommunities are defined to be collections of ASVs that co-occur in samples at given relative proportions.  An ASV can occur in multiple subcommunities at various abundances and the key assumption underlying an MM model, in contrast to a clustering model, is that different instances (i.e., different sequencing reads) of the same ASV in a sample can arise from the participation of that ASV in multiple microbial subcommunities. Key parameters of interest in MM models are subcommunity abundance, i.e., the proportions of the various subcommunities in each sample, and subcommunity composition, i.e., the proportions of the ASVs in each subcommunity.

To describe LDA, it is convenient to introduce categorical indicators for each read and its associated subcommunity identity. For $d=1,2,\ldots,D$, let $\w_d$ be a vector $\w_d = (w_{d,1},\dots,w_{d,N_d})$ where $w_{d,n}\in \{1,2,\ldots,V\}$ is the categorical indicator of the ASV associated with the $n$th read in the sample.  We refer to the elements $w_{d,n}$ in this vector as ``tokens" to draw analogy with topic modelling.  There, each token is a word in a document; here, each token corresponds to a read in a sample. We also note that $x_{d,v}=\sum_{n=1}^{N_d} \mathbf{1}_{\{w_{d,n}=v\}}$.
  
Let $\bm{\phi}_d=(\phi_{d}^1,\phi_{d}^2,\ldots,\phi_{d}^K)' \in \Delta^{K-1}$, where $\Delta^{S}$ is the $S$-dimensional simplex, be the subcommunity abundance vector. That is, $\phi_{d}^k$ represents the relative abundance of subcommunity $k$ in sample $d$, and so $\bm{\phi}_d$ specifies the categorical distribution of each token over the $K$ underlying subcommunities in sample $d$. Let $z_{d,n}$ represent the subcommunity from which the $n^{th}$ token in sample $d$ arises from and let $\z_d$ be the vector all such assignments for sample $d$.  Also, let $\bm{\beta}_k=(\beta_{k}^1,\beta_{k}^2,\ldots,\beta_{k}^V)' \in \Delta^{V-1}$ be the subcommunity composition for subcommunity~$k$. That is, $\bm{\beta}_k$ gives the relative proportions of the $V$ unique ASVs in subcommunity $k$. For $d=1,\dots,D$ and $n=1,\dots,N_d$ and while $\bm{\alpha}$ and  $\bm{\gamma}$ are hyperparameters, the LDA model (Figure~$\ref{fig:graph_lda}$) \citep{2003_Blei} is then \\
\[
\begin{aligned}
w_{d,n}\,|\,z_{d,n},\bm{\beta}_{z_{d,n}} &\ind {\rm Cat}(\bm{\beta}_{z_{d,n}}) &
z_{d,n}\,|\,\bm{\phi}_d &\ind {\rm Cat}(\bm{\phi}_d) \\
\bm{\phi}_d\,|\,\bm{\alpha} &\iid {\rm Dir}(\bm{\alpha})&
\bm{\beta}_k\,|\,\bm{\gamma} &\iid {\rm Dir}(\bm{\gamma}).
\end{aligned}
\]

Though LDA can be applied in the microbiome context \citep{2019_Sankaran}, it does not account for cross-sample heterogeneity in subcommunity composition. In particular, it assumes that the $\bm{\beta}_k$ are the {\em exact same} across all samples.  This is inconsistent with the empirical behavior of the microbiome where large cross-sample heterogeneities exist \citep{2012_Holmes}.  LDA thus tends to interpret cross-sample heterogeneity as the presence of additional subcommunities.

\subsection{Incorporating cross-sample heterogeneity}

We shall enrich the LDA framework to allow the subcommunity compositions to vary across samples.  There are several hierarchical models for microbiome compositions such as the Dirichlet-Multinomial (DM) model \citep{2000_Nigam,2012_Holmes} and Aitchinson's log-ratio based normal (LN) models \citep{1982_Aitchison}, which could be embedded into LDA for this purpose. However, the DM is highly restrictive in its ability to characterize the underlying cross-sample variability as the Dirichlet distribution has only one scalar variance parameter, while the LN models are computationally challenging due to lack of conjugacy to the multinomial sampling model. To resolve these difficulties, we adopt the recently introduced logistic-tree normal (LTN) model \citep{2021_Wang}. In particular, we will show that the LTN model can be embedded into the LDA model to accommodate cross-sample heterogeneity and that posterior inference can be accomplished through simple collapsed blocked Gibbs sampling using a data-augmentation technique called P\'olya-Gamma augmentation.  Moreover, since the adoption of the LTN model requires specifying a dyadic partition tree on the ASVs, the phylogenetic tree relating the taxa is a natural choice. 

\subsubsection{The phylogenetic tree}

Let $\mathcal{T}$ denote a phylogenetic tree capturing genetic similarities between the observed ASVs. The leaf nodes in the tree correspond to the observed ASVs in the data set.  Each interior node is the inferred common ancestral taxon for the ASVs lying in the corresponding descendant subtree at the node. Each node (or taxon) $A$ in the phylogenetic tree $\mathcal{T}$ can be represented by the collection of its descendant ASVs. In particular, each leaf node $A$ contains a single ASV, whereas each internal node $A$ contains multiple ASVs. In the following, we let $\mathcal{I}$ be the set of internal nodes. Throughout this work, we shall assume that the phylogenetic tree is rooted and binary in the sense that each $A\in\mathcal{I}$ has exactly two child nodes (i.e., direct descendants): let $A_l$ and $A_r$ be the left and right children of $A$, respectively.

\subsubsection{The logistic-tree normal model}

\begin{figure}
\begin{subfigure}{.45\textwidth}
  \centering
  \includegraphics[width=0.95\linewidth]{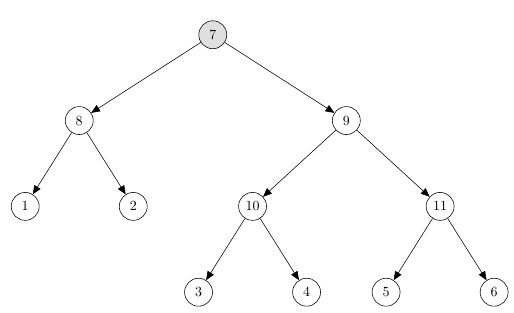}
  \caption{Notation for nodes}
  \label{fig:toy_tree}
\end{subfigure}%
\begin{subfigure}{0.45\textwidth}
  \centering
  \includegraphics[width=0.95\linewidth]{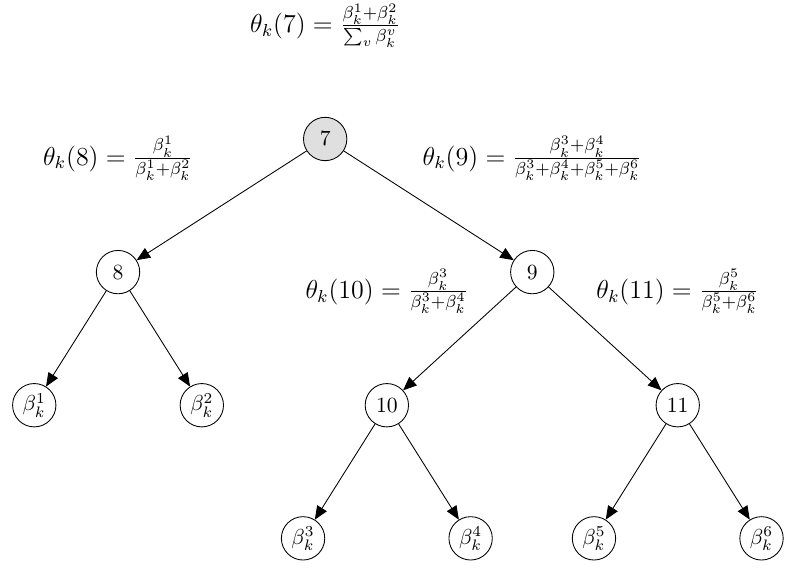}
  \caption{Present $\theta_k$ and $\beta_k$}
  \label{fig:labelled_toy_tree}
\end{subfigure}
\caption{An example phylogenetic tree for $6$ ASVs and the graphical relationship between $\mu_k$ and $\psi_{d,k}$.}
\label{fig:toy_trees}
\end{figure}

We shall adopt the logistic-tree normal (LTN) model \citep{2021_Wang} as the sampling model for the ASV count distribution within each subcommunity. LTN is a distribution on a tree-based log-odds transform of the categorical probabilities $\bm{\beta} = (\beta^1,\beta^2,\ldots,\beta^V)' \in \Delta^{V-1}$. Specifically, given the phylogenetic tree $\mathcal{T}$, for each interior node we define $\theta(A) = \frac{\sum_{v\in A_l}\beta^v}{\sum_{v\in A}\beta^v}$: the probability that a token belongs to an ASV in $A_l$ given that it belongs to an ASV in $A$. The collection of $\theta(A)$ on all $A\in\mathcal{I}$ gives an equivalent reparametrization of $\bm{\beta}$. In Figure~$\ref{fig:toy_trees}$ we plot an example phylogenetic tree over $6$ ASVs with labelled nodes (Figure~$\ref{fig:toy_tree}$) and with labelled $\beta^v$ and $\theta(A)$ (Figigure~$\ref{fig:labelled_toy_tree}$) to demonstrate the link between the $\beta^v$ and the $\theta(A)$. 

After taking the logit transform of these binomial probabilities on the tree nodes, $\psi(A) = \log \frac{\theta(A)}{1-\theta(A)}$, let $\bm{\psi}$ be the vector of $\psi(A)$ with respect to an ordering on the $p$ internal nodes of $\mathcal{T}$.  LTN is simply a Gaussian model on these tree-based log-odds: $\bm{\psi} \, | \, \bm{\mu},\Sigma \iid {\rm MVN}(\bm{\mu},\Sigma)$ for some mean $\bm{\mu}$ and covariance $\Sigma$ parameters that specify the overall average profile of the count distribution and the cross-sample variability.

Posterior computation under LTN, which we will describe later, relies on an equivalent representation of the categorical sampling  on the leaves of the tree as a collection of sequential binomial experiments on the internal nodes of the tree. Specifically, generating a categorical draw from the probability vector $\bm{\beta}$ can be achieved by sequentially ``dropping'' the token from top-to-bottom along the phylogenetic tree: at each node determine whether the token belongs to the left or right child node with probabilities $\theta(A)$ and $1-\theta(A)$, respectively. More formally, for each node $A\in\mathcal{T}$, we use $y(A)$ to denote the total counts associated with the ASVs descended from node $A$. That is, $y(A) = \sum_{n=1}^{N} 1_{w_{n}\in A}$ where $w_n$ represents the $n$th count.  Generating a multinomial count vector with probabability $\bm{\beta}$ can be achieved by sequentially drawing  $y(A_l)$ given $y(A)$ from $\text{Bin}(y(A_l)\,|\,y(A),\theta(A))$. Putting the pieces together, and letting ${\rm expit}(\psi)=1/(1+e^{-\psi})$, LTN is the following generative model: for all internal nodes $A\in \mathcal{T}$,
\[
\begin{aligned}
y(A_l)\,|\,y(A),\psi(A) &\ind {\rm Bin}(y(A),\theta(A)={\rm expit}(\psi(A))) & \text{ and } \quad 
\bm{\psi}\,|\,\bm{\mu},\Sigma &\ind {\rm MVN}(\bm{\mu},\Sigma).
\end{aligned}
\]

\subsection{LTN-LDA}

We incorporate the LTN model into LDA to allow cross-sample heterogeneity in subcommunity compositions. The resulting model is termed logistic-tree normal latent Dirichlet allocation (LTN-LDA). Specifically, for $d = 1,\dots,D$, $k=1,\dots,K$, $n=1,\dots,N_d$, and $A\in \mathcal{I}$, where the subscripts $d$, $k$, and $n$ indicate the corresponding quantities associated with the $d$th sample, $k$th subcommunity, and $n$th read, the model is as follows \\
\[
\begin{aligned}
y_{d,k}(A_l)\,|\,y_{d,k}(A),\psi_{d,k}(A) &\ind {\rm Bin}(y_{d,k}(A),\text{expit}(\psi_{d,k}(A))) \\
y_{d,k}(A) & =\sum_{n=1}^{N_d} 1_{z_{d,n}=k}1_{w_{d,n}\in A} &
z_{d,n}\,|\,\bm{\phi}_d &\ind {\rm Cat}(\bm{\phi}_d)\\
\bm{\phi}_d \,|\, \bm{\alpha} &\overset{\text{iid}}{\sim} {\rm Dir}(\bm{\alpha})&
\bm{\psi}_{d,k}\,|\,\bm{\mu}_k,\Sigma_k &\ind {\rm MVN}(\bm{\mu}_k,\Sigma_k), \\
\bm{\mu}_k \,|\, \bm{\mu}_0, \Lambda_0 &\overset{\text{iid}}{\sim} {\rm MVN}(\bm{\mu}_0,\Lambda_0) &
\Sigma_k \,|\, {\rm G} &\overset{\text{iid}}{\sim} {\rm G}
\end{aligned}
\]
Note that we also endowed the subcommunity mean $\bm{\mu}_k$ and covariance $\Sigma_k$, with corresponding priors ${\rm MVN}(\bm{\mu}_0,\Lambda_0)$ and $G$, which will be specified later.  Figure~$\ref{fig:graph_ltn_lda}$ provides the graphical model representation for this full hierarchical model. The key distinction between LTN-LDA and LDA is that LTN-LDA uses a hierarchical kernel, namely LTN,  to model cross-sample heterogeneity. In particular, the composition  in sample $d$ of subcommunity $k$ is determined by $\bm{\psi}_{d,k}$ and is  explicitly allowed to vary across samples. 

Without additional constraints on the high-dimensional covariance matrices for each subcommunity, $\Sigma_k$, the model is too flexible \citep{2009HaffariTeh}, and can become unidentifiable. Additional structural constraints serving the purpose of regularization on the covariance structure are thus necessary and so we assume that $\Sigma_k$ is a diagonal covariance matrix.  An LTN distribution with diagonal covariance is similar in distributional properties to a  Dirichlet-tree multinomial (DTM) distribution \citep{1991Dennis,2017_Wang} but is computationally more efficient because there are no known conjugate priors for the mean and variance parameters under the DTM model. While this limitation is manageable when the DTM is used as a standalone model or the top layer in a hierarchical model, when embedded as a kernel within an MM model such as LDA the incurred numerical computational cost becomes prohibitive. (For more details, see Supporting Information S1.)

While the covariance constraint may appear strong, we note that the dependence among the tree-based log-odds ratios is generally much weaker than the complex dependence structure among the ASV counts themselves. In a sense, the tree-based log-odd transform of the abundance vectors ``decorrelates'' the data. For the interested reader, this decorrelation phenomenon is analogous to the so-called ``whitening'' effects in wavelet analysis \citep{2008Nason}, as the dyadic tree transform we incorporate here is the counterpart of Haar-wavelet transform on functions.  (In Supporting Information Section~S2 we investigate the effects of relaxing the diagonal covariance to a blocked diagonal covariance, and the results show that the additional sophistication does not lead to noticeable improvement in the inference.)

Aside from the diagonal covariance, we also assume that the amount of variability for each node depends on that node's distance to the bottom (i.e., leaf) level of the tree. In particular, we assume that taxa close to the bottom of the phylogenetic tree have larger cross-sample variability in the corresponding log-odds ratio than those which are distant. This is motivated by the biological intuition that taxa close to each other on deep levels of the phylogenetic tree tend to have comparable functionality; the relative proportions of such taxa thus often display elevated levels of variance  \citep{2021JeganathanHolmes}. 

Specifically, let $|A|$ measure the distance of $A$ from the leaf level by denoting the number of leaves descended from node $A$. For $i = 1,\dots,p$, $k = 1,\dots, K$, $C\in\mathbb{N}$ (a tuning parameter), and $\bm{\tau}_k = (\tau_{k}^1,\dots,\tau_{k}^p)$, the prior we adopt has the form $\Sigma_k \,|\, \bm{\tau}_k = \text{diag}(\bm{\tau_{k}})$ where
\[
\tau_{k}^i\,|\,a_1,a_2,b \iid
\begin{cases}
\text{IG}(a_1,b) & |A_i| \geq C \\
\text{IG}(a_2,b) &|A_i| < C \\
\end{cases}
\]
We default to $(a_1,a_2,b) = (10^4,10,10)$ and note that while we still refer to the $\psi_{d,k}$ as being drawn from a multivariate normal distribution, we have $\psi_{d,k}^i \, | \, \mu_k^i,\tau_k^i \iid \text{N}(\mu_k^i,\tau_k^i)$.

This choice of priors ensures conjugate updating and avoids identifiability issues. Further, it partitions the internal nodes of the tree in two: we shall refer to these sets as the upper tree $\mathcal{U} = \{A\in\mathcal{I}:|A|\geq C\}$ and the lower tree $\mathcal{L} =\{A\in\mathcal{I}:|A| < C\}$.  In $\mathcal{U}$, the hyperparameters $a_1$ and $b$ are such that the $\tau_{k}^i$ will be small and the $\psi_{d,k}^i$ will vary little around $\mu_k^i$; in $\mathcal{L}$, the hyperparameters $a_2$ and $b$ are such that the $\tau_k^i$ are allowed to be large and the $\psi_{d,k}^i$ can vary significantly across samples.  This implies that if $A_c$ is the child of $A$, and $A_c\in\mathcal{L}$ but $A\in\mathcal{U}$, then all ASVs descended from $A_c$ can substitute for each other across samples in a given subcommunity.  We call sets of ASVs which are allowed to substitute for each other substitution sets.  All ASVs are either part of a substitution set or singletons.  The tree structure is critical to how LTN-LDA models cross-sample heterogeneity, and we include an analysis on the robustness to misspecified trees in Supporting Information S3.

\subsection{Bayesian inference by collapsed blocked Gibbs sampling}

\begin{figure}
\begin{subfigure}{.2\textwidth}
  \centering
  \includegraphics[width=0.95\linewidth]{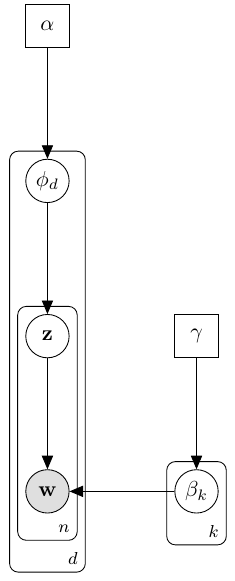}
  \caption{LDA}
  \label{fig:graph_lda}
\end{subfigure}%
\begin{subfigure}{0.4\textwidth}
  \centering
  \includegraphics[width=0.95\linewidth]{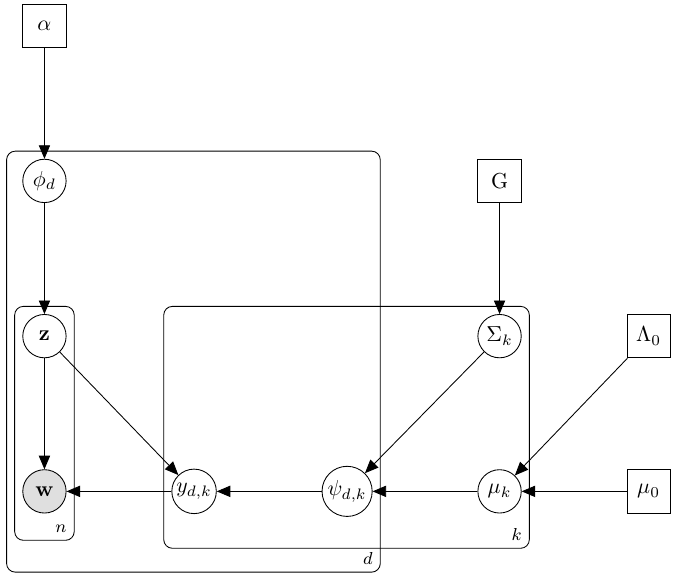}
  \caption{LTN-LDA}
  \label{fig:graph_ltn_lda}
\end{subfigure}
\begin{subfigure}{.4\textwidth}
  \centering
  \includegraphics[width=0.95\linewidth]{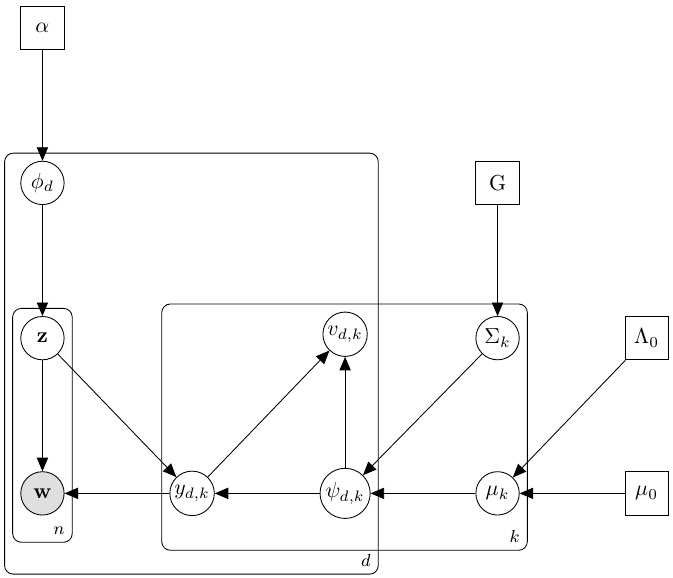}
  \caption{LTN-LDA with P\'olya-Gamma Variables}
  \label{fig:graph_ltn_lda_pg}
\end{subfigure}
\caption{Graphical model representations for LDA and LTN-LDA.}
\label{fig:graph_models}
\end{figure}

While the LTN-LDA model  is not conditionally conjugate by itself, one can restore conjugacy by introducing a class of P\'olya-Gamma latent variables \citep{2013_Polson} $v_{d,k}(A)$ --- one for each interior node $A$ --- which are independent of $y_{d,k}(A_l)$ conditioned on $y_{d,k}(A)$ and $\psi_{d,k}(A)$: $v_{d,k}(A)\,|\,y_{d,k}(A),\psi_{d,k}(A) \sim {\rm PG}(y_{d,k}(A),\psi_{d,k}(A))$. The full conditional for $\psi_{d,k}(A)$ is then proportional to $\exp\left({(y_{d,k}(A_l) - \frac{y_{d,k}(A)}{2})\psi_{d,k}(A) - \frac{v_{d,k}(A)\psi_{d,k}(A)^2}{2}}\right)$,
which takes a quadratic form in the exponent and thus is conjugate to the Gaussian model on $\psi_{d,k}(A)$. The graphical model for LTN-LDA with the P\'olya-Gamma variables is presented in Figure $\ref{fig:graph_ltn_lda_pg}$.  To speed up the sampling of P\'olya-Gamma variables we adopt an approximate sampler proposed by \cite{2019_Glynn} for $y_{d,k}(A) \geq 30$.  Further, we integrate $\bm{\phi}_d$ out of the sampling model to improve convergence as in \citep{2004_Griffiths}.  The algorithm scales linearly with $D$, $K$, $V$, and $N_d$.  (For details, see Supporting Information S4.)

\section{Numerical experiments}

\subsection{Robustness in choosing the number of subcommunities}
The true number of subcommunities $K$ in a given dataset is typically unknown and it is common to treat $K$ as a tuning parameter. However, for data with large cross-sample heterogeneity such as microbiome data, intuition suggests that a model assuming zero heterogeneity will confuse sample-specific variation around a subcommunity mean with the presence of additional subcommunities.  This results in difficulty estimating $K$ and inference sensitive to $K$; indeed, LDA encounters both of these difficulties \citep{2021Fukuyamaetal}.

\begin{figure}
    \centering
    \includegraphics[width = \textwidth]{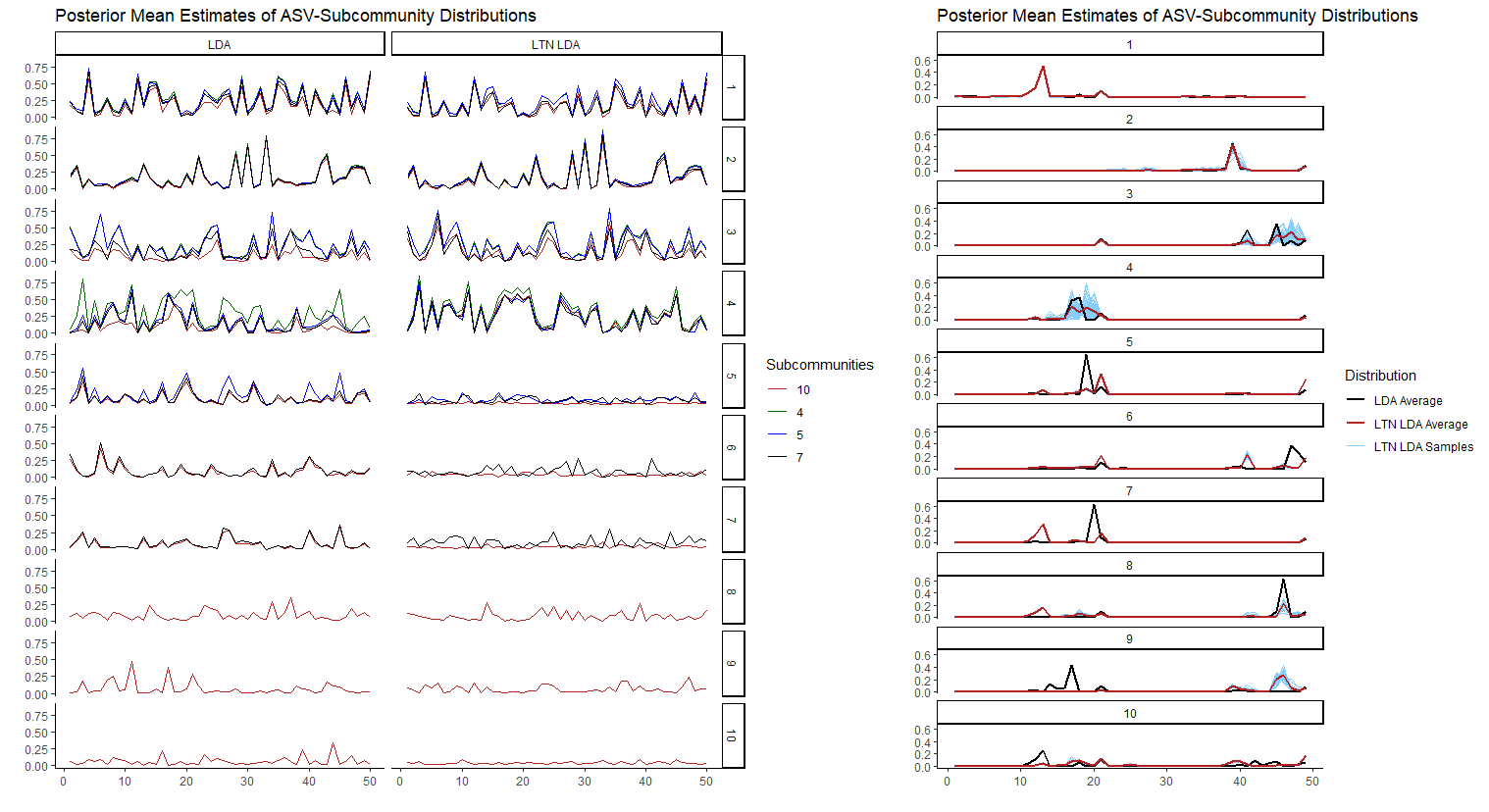}
    \caption{(Left) Subcommunity abundance for $\bm{\phi}_d$ for all samples over four different numbers of subcommunities $K \in \{4,5,7,10\}$ for LDA (left) and LTN-LDA (right). The estimated abundances are noticeably more stable over different values of $K$ for the LTN-LDA.  (Right) Estimated subcommunity compositions for all samples. Blue indicates the sample-specific composition under LTN-LDA ($\bm{\beta}_{d,k}$, red indicates the average subcommunity composition under LTN-LDA ($\bm{\beta}_{k}$) and black indicates the average subcommunity composition ($\bm{\beta}_{k}$) under LDA.  The 49  ASVs are on the x-axis.}
    \label{fig:Sim_Case_K_10_Mult_Plots}
\end{figure}

To verify this intuition, we generated data from a known LTN-LDA model which induces cross-sample heterogeneity. In particular, we simulated $D = 50$ samples, and $N_d = 10,000$ reads per sample; we set $\alpha = 1$, $\mu = 0$, $\Lambda = I$, $a_1 = 10^4$, $a_2=b=10$, and $(K,C) = (4,5)$.  The underlying phylogenetic tree is presented in Supporting Information S5: there are $V = 49$ ASVs.  We then contrasted LDA and LTN-LDA by running Gibbs samplers on the data generated above with $K\in\{4,5,7,10\}$ and $C = 5$. In the left part of Figure~$\ref{fig:Sim_Case_K_10_Mult_Plots}$, we plot the posterior means of the subcommunity abundances $\bm{\phi}_d$ for both LDA and LTN-LDA. We corrected for label switching and estimated the $\bm{\phi}_{d}$ as in \cite{2004_Griffiths}.

With $K$ set to truth, LDA performs comparably to LTN-LDA in estimating the true values of $\bm{\phi}_d$; however, as we increase $K$, the inference provided by LDA worsens.  While it still recovers the abundances for subcommunities $1$ and $2$, it does a worse job at recovering subcomunities $3$ and $4$.  Moreover, LDA detects the presence of additional subcommunities which do not exist in the true generative model.  LTN-LDA, in contrast, is remarkably stable when $K$ is overspecified.  No matter the modelled value of $K$, it detects the four true subcommunities with approximately the same abundances while estimating that additional subcommunities have little abundance.  For $K = 10$, we plotted the subcommunity compositions on the right part of Figure~\ref{fig:Sim_Case_K_10_Mult_Plots}.  For LTN-LDA, distributions for the $\bm{\beta}_{d,k}$ are in blue and the $\bm{\beta}_k$ are in red; the LDA $\bm{\beta}_k$ distributions are in black.  LTN-LDA finds moderate levels of cross-sample heterogeneity in subcommunity $2$, and a high levels in samples $3$ and $4$.  

These figures imply that LDA is able to recover the subcommunity abundances only for those subcommunities with low cross-sample heterogeneity. LDA fails to recover the subcommunity abundances for those subcommunities with high cross-sample heterogeneity, mistaking heterogeneity for additional subcommunities.  In effect, LDA splits true heterogeneous subcommunities into many smaller subcommunities with no heterogeneity and ASVs which ought to belong in the same subcommunity are separated.  LTN-LDA, on the other hand, provides stable and accurate inference as the modelled $K$ increases.  This thus confirms our intuition about the behavior of LDA in the presence of cross-sample heterogeneity.

\subsection{Predictive scoring as a device for choosing tuning parameters}

While incorporating cross-sample heterogeneity enhances the robustness of LTN-LDA to overspecifying the number of subcommunities, it is still useful to have a generally applicable strategy for setting the tuning parameters for LTN-LDA: $K$ and $C$. One option is to use out-of-sample predictive performance to identify suitable choices of the tuning parameters. A popular performance measure for MM models is perplexity \citep{2009_Wallach}: a transform of out-of-sample predictive likelihood such that lower perplexity is preferred.

We thus implement the simple strategy of computing the average out-of-sample perplexity score for different choices of $(K,C)$ and examine whether that can lead to a practical way of choosing these parameters. We will also examine whether this strategy could be adopted for models without cross-sample heterogeneity, namely LDA, to alleviate their limitations. We follow the procedure in Section~$5.1$ of \cite{2009_Wallach} for computing the perplexity for LDA, and generalize that strategy to LTN-LDA.  (For deatils, see Supporting Information S6.)  We generated 200 simulated datasets.  In each, there are $D = 50$ samples and $N_d = 10,000$ counts per sample; we set $\alpha = 1$, $\mu = 0$, $\Lambda = I$, $a_1 = 10^4$, $a_2=b=10$, and $(K,C) = (4,5)$.  For each dataset, we also generate a test set of the same size where the sample specific parameters are generated using $\alpha = 1$ and the training set's $\bm{\mu}_k$ and $\Sigma_k$.

\begin{figure} [H]
\centering
\begin{tabular}{cccc}
\includegraphics[width=.3\textwidth]{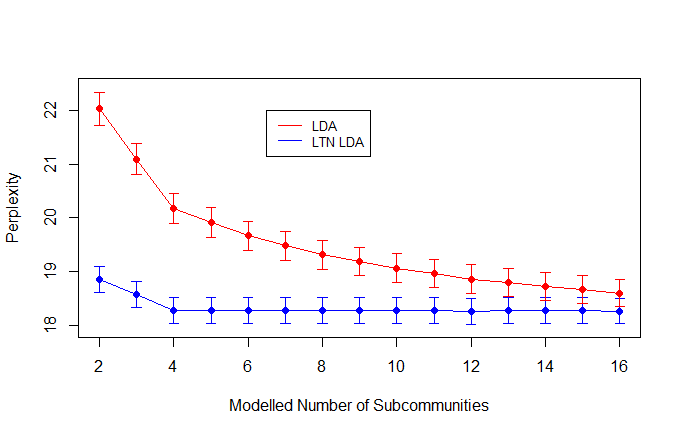} &
\includegraphics[width=.3\textwidth]{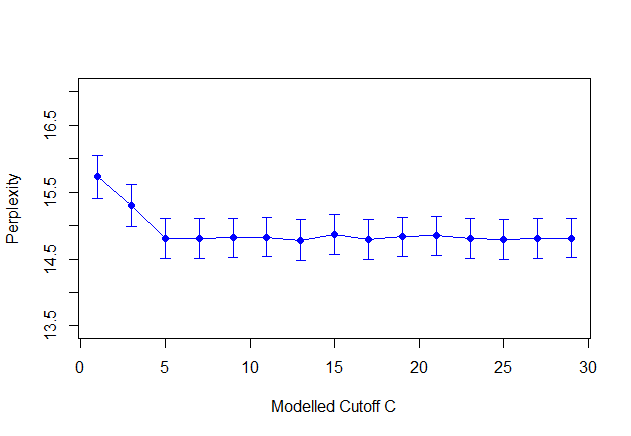} &
\includegraphics[width=.3\textwidth]{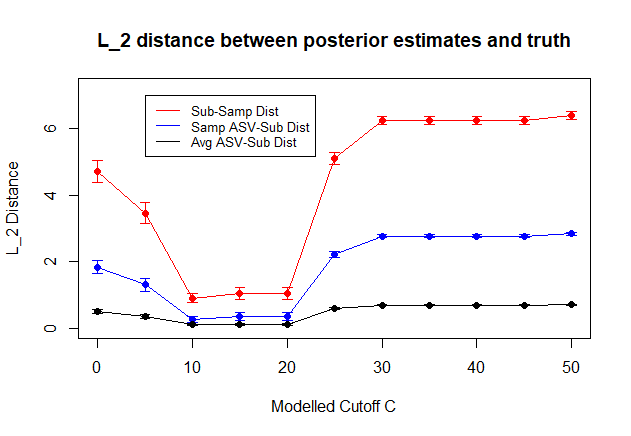} \\
\textbf{(a)} & \textbf{(b)} & \textbf{(c)}  \\[6pt]
\end{tabular}
\begin{tabular}{cccc}
\includegraphics[width=.3\textwidth]{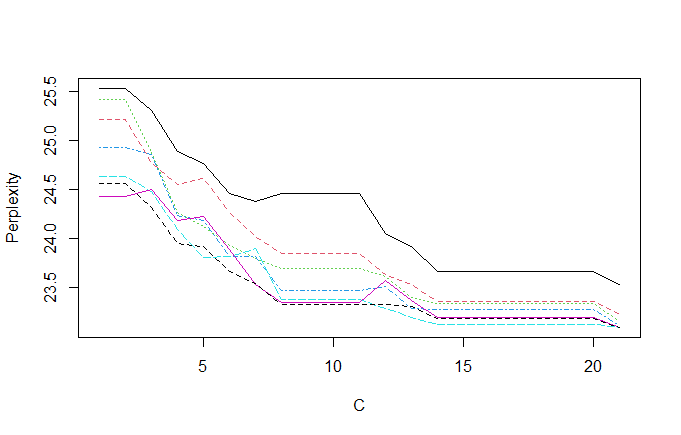} &
\includegraphics[width=.3\textwidth]{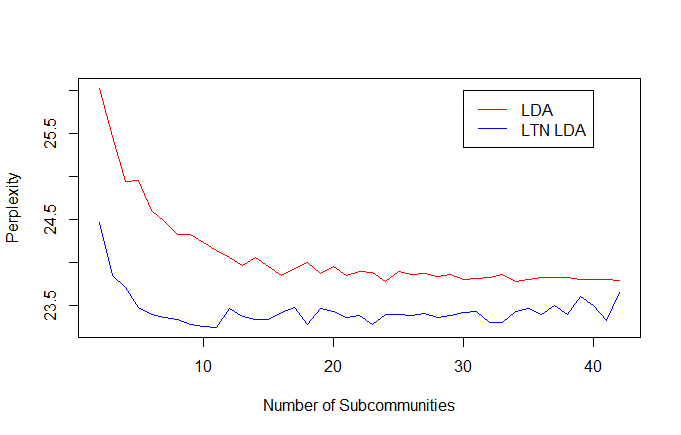} \\
\textbf{(d)}  & \textbf{(e)}  \\[6pt]
\end{tabular}
\caption{ \textbf{(a)} Perplexity for LDA (red) and LTN-LDA (blue) as the modelled number of subcommunities $K$ varies. 
\textbf{(b)} Perplexity for LTN-LDA as the modelled threshold $C$ varies.
\textbf{(c)} $L_2$ distances for $\bm{\phi}_d$ (red), $\bm{\beta}_{k,d}$ (blue), and $\bm{\beta}_{k}$ (black) for LTN-LDA as the modelled threshold $C$ varies.
\textbf{(d)} Perplexity for varying levels on $K$ on the Dethlefsen and Relman data as we vary $C$.
\textbf{(e)} Perplexity for LTN-LDA and LDA as $K$ varies while $C = 8$.}
\label{fig:Perp_Results}
\end{figure}

Fixing $C$ to truth, we varied $K$ and computed average perplexity for LDA and LTN-LDA in Figure~\ref{fig:Perp_Results}(a).  There are three main observations: (i) LTN-LDA significantly outperforms LDA for $K$ near truth, (ii) the perplexity curve for LTN-LDA decreases until it stabilizes at the true value of $K$, (iii) the perplexity curve for LDA continues to decrease as the modelled $K$ is increased past its true value.  The main reason for the difference is that LDA interprets the presence of cross-sample heterogeneity as extra subcommunities and so finds as many subcommunities as are modelled.  While this improves out-of-sample predictive performance, it does not improve inference on the underlying truth.  Thus, using perplexity to select the modelled number of subcommunities for LDA is a poor method if there is significant cross-sample heterogeneity.  LTN-LDA is more robust and parsimonious in its representation of the data because it incorporates cross-sample heterogeneity in subcommunity compositions.

Fixing $K$ to truth, we computed average perplexity for LTN-LDA as we varied $C$ in Figure~\ref{fig:Perp_Results}(b).  The perplexity curve decreases until it stabilizes at the true value of $C$.  In addition to perplexity, we also computed the $L_2$ distances between the posterior mean estimates and the true values for the $\bm{\phi}_d$, $\bm{\beta}_{d,k}$, and $\bm{\beta}_k$ distributions (Figure~\ref{fig:Perp_Results}(c)).  Unlike the perplexity curves, the $L_2$ distances are lowest around $C = 5$ and increase as $C$ increases.  Thus, if the modelled value of $C$ is increased too far above truth, inference becomes unreliable.  

The above results suggest a simple two-stage strategy for choosing $(K,C)$ using perplexity. First, let $(K,C)$ vary jointly on a grid and use cross-validation to compute the average perplexity, giving $K$ perplexity curves over $C$.  Set $C$ to be the inflection point in these curves. Second, vary $K$ and set the value of $K$ to be the inflection point of the resulting perplexity curve. Note that this strategy may fail for LDA: as our numerical examples show below, due to the lack of cross-sample heterogeneity in LDA, the perplexity score generally continues to improve as one increases the number of subcommunities beyond truth.  This in turn leads to misleading inference on subcommunity abundance and composition.

\section{Evaluation on a microbiome study}

We apply LTN-LDA to identify subcommunity dynamics in the dataset of \cite{2011_Dethlefsen}, which has been previously investigated by \cite{2019_Sankaran} using LDA.  The data includes gut microbiome samples of three patients who were administered two five-day courses of ciprofloxacin  over a ten-month span.  We focus on the $54$ samples from patient F, each consisting of approximately $10,000$ reads. Ciproflaxin was administered during samples 12-23 and 41-51. There are $2,852$ unique ASVs in the dataset; we merged ASVs into taxa at the finest known level and pruned all taxa which did not total at least $100$ sequencing reads.  This left $44$ taxa comprising $99.86$ percent of the original counts.  The resulting phylogenetic tree is included in Supporting Information~S7.  

We implemented the strategy outlined above to choose tuning parameters.  In particular, we implement a $4$-fold cross-validation letting $K$ vary in $\{2,3,\dots,8\}$ and $C$ in $\{1,2,\dots,21\}$.  The resulting $K$ perplexity curves over $C$ are presented in Figure $\ref{fig:Perp_Results}$(d).   The inflection point in the curve appears at $C = 8$.  Setting $C = 8$ and varying $K$ gives the results in Figure $\ref{fig:Perp_Results}$(e); for comparison, we also applied LDA to the data over varying $K$. LTN-LDA has strictly lower perplexity than LDA, indicating that there are significant levels of cross-sample heterogeneity in the dataset. Moreover, LTN-LDA experiences a noticeable inflection point (near $K = 5$) in contrast to LDA whose perplexity decays slowly. 

\begin{figure}
    \centering
    \includegraphics[width = \textwidth]{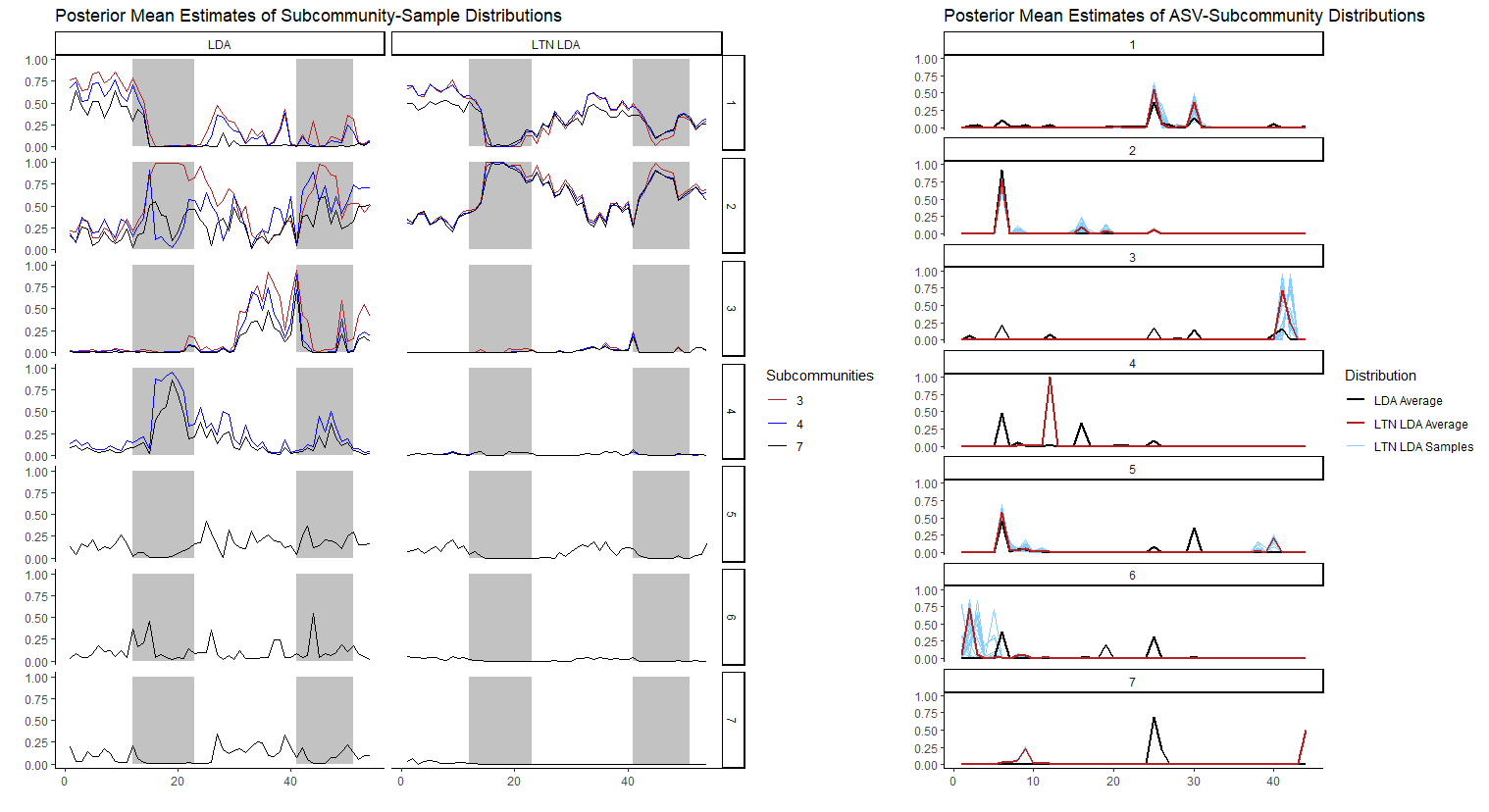}
    \caption{(Left) Subcommunity abundance for $\bm{\phi}_d$ for all samples over three different numbers of subcommunities $K \in \{3,4,7\}$ for LDA (left) and LTN-LDA (right). The estimated abundances are noticeably more stable over different values of $K$ for the LTN-LDA.  (Right) Estimated subcommunity compositions for all samples. Blue indicates the sample-specific composition under LTN-LDA ($\bm{\beta}_{k,d}$, red indicates the average subcommunity composition under LTN-LDA ($\bm{\beta}_{k}$) and black indicates the average subcommunity composition ($\bm{\beta}_{k}$) under LDA.  The 44  ASVs are on the x-axis.}
    \label{fig:Sank_C_8_K_7_Mult_Plots}
\end{figure}

\begin{figure}
    \centering
    \includegraphics[width = \textwidth]{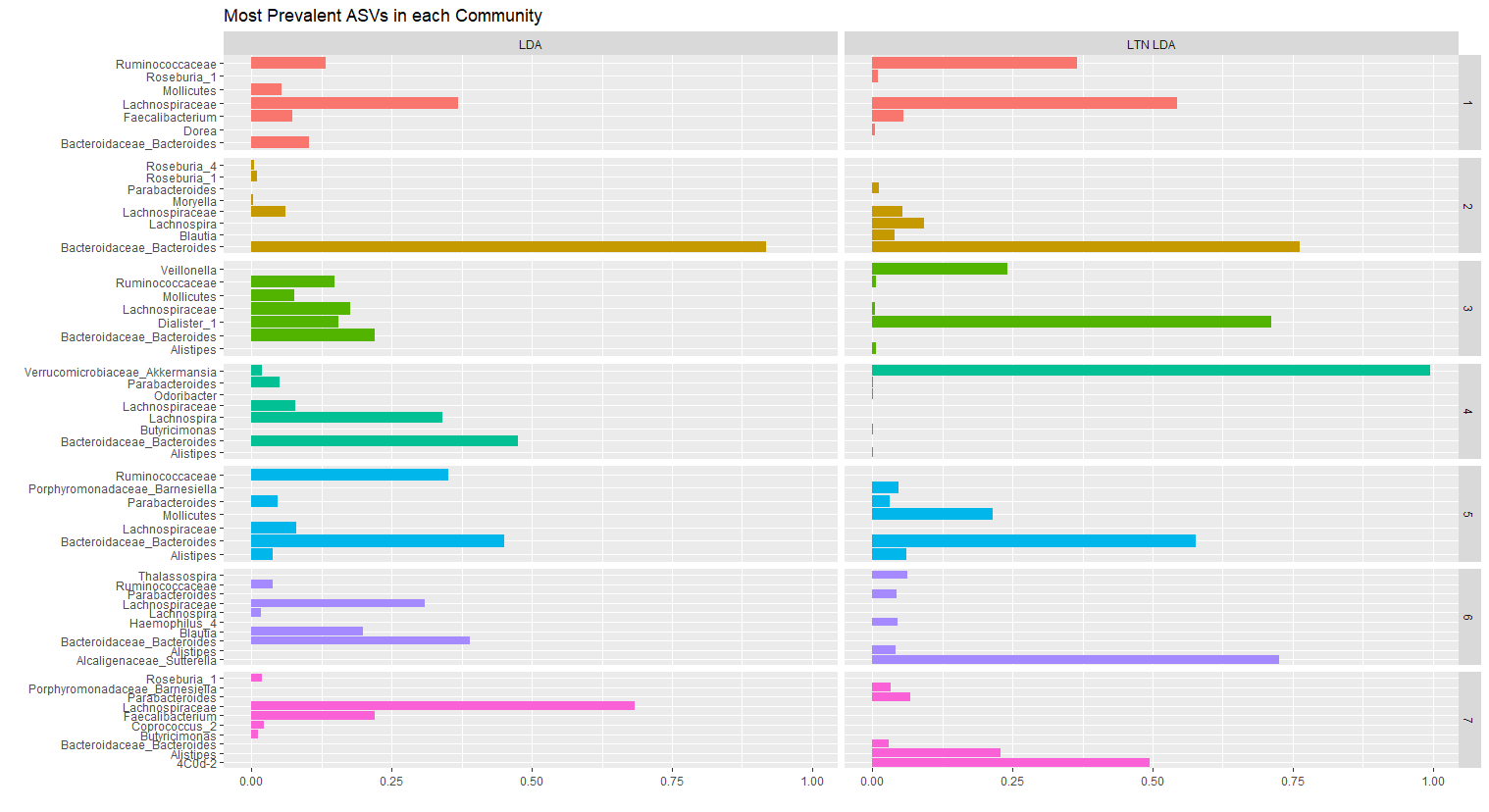}
    \caption{The $5$ most prevalent ASVs in each subcommunity for LDA and LTN-LDA, $K = 7$, $C = 8$.}
    \label{fig:Sank_C_8_K_7_Top}
\end{figure}

We now present more detailed analysis for LTN-LDA and LDA with $C = 8$.  For $K \in \{3,4,7\}$ we plotted the subcommunity abundance on the left side of Figure $\ref{fig:Sank_C_8_K_7_Mult_Plots}$, after manually correcting for label switching. The grey regions indicate periods of ciproflaxin treament.  The subcommunities found by LTN-LDA are remarkably stable as $K$ changes.  Subcommunities $1$, $2$, and $3$ have almost the exact same abundance, and additional subcommunities have minimal abundance.  LDA, however, finds as many subcommunities as are modelled: it will split a heterogenous subcommunitiy into multiple subcommunities with no heterogeneity.  For $K = 7$, we plotted the ASV-subcommunity distributions on the right side of Figure $\ref{fig:Sank_C_8_K_7_Mult_Plots}$.  Distributions for the $\bm{\beta}_{d,k}$ are in blue, the $\bm{\beta}_k$ in red, and the LDA distributions in black.  The $5$ most prevalent ASVs in each subcommunity are presented in Figure $\ref{fig:Sank_C_8_K_7_Top}$ for LDA and LTN-LDA.  These demonstrate that LTN-LDA finds significant levels of cross-sample heterogeneity and subcommunities with meaningfully different compositions than LDA.  

LTN-LDA thus provides two major advantages.  First, LTN-LDA is more robust with respect to modelling differing numbers of subcommunities than LDA.  This is similar to our simulations and indicates that LTN-LDA better accounts for the cross-sample heterogeneity in the data than does LDA.  Moreover, the three subcommunities found by LTN-LDA are biologically interpretable.  The first subcommunity is composed mostly of Lachnospiraceae and Ruminococcaceae and displays significant levels of cross-sample heterogeneity, indicating that LTN-LDA has found these two ASVs can substitute for each other.  \cite{2018Haaketal} found this phenomena in humans undergoing ciproflaxin treatment.  LTN-LDA can thus learn when two ASVs substitute for each other across samples from the data, with no prior knowledge. The second subcommunity, composed mainly of Bacteroides, increases in abundance during the antibiotic treatments.  Studies in mice \citep{2020_Zhu} and humans \citep{2015_Stewardson} indicate that the abundance of Bacteroides increases during ciproflaxcin treatment.  The third subcommunity has a small spike in abundance only on the first day of the second antibiotic course, and is composed mostly of Dialister and Veillonella.  Ciproflaxin has been shown to be effective against Dialister \citep{2007_Morio} which may explain the decrease in this subcommunity after treatment began. 

\section{Discussion}

We have proposed a novel mixed-membership model which seeks to appropriately incorporate cross-sample heterogeneity in subcommunity compositions: a characteristic of the data prevalent in most microbiome studies.  By incorporating the logistic-tree normal model for the sample-specific compositions of each subcommunity, we explicitly allow the composition of subcommunities to vary across samples.  We have shown that incorporating cross-sample heterogeneity into MM models can lead to substantially improved inference over models which assume zero cross-sample heterogeneity.  LTN-LDA is substantially more robust than LDA with respect to overspecifying $K$ and significantly outperforms LDA in terms of predictive performance.  Moreover, perplexity can be a useful device to set the tuning parameters for LTN-LDA but not for LDA.  Posterior computation on LTN-LDA can proceed through collapsed blocked Gibbs-sampling with the assistance of P\'olya-Gamma augmentation, and as such implementation for LTN-LDA is convenient.  Moreover, LTN-LDA is a fully Bayesian model and the Gibbs sampler allows for posterior uncertainty quantification.

In comparison to LDA, LTN-LDA incorporates two new features: the tree structure and the random effects allowing cross-sample heterogeneity.  The tree structures provides guidance on how to parsimoniously model the random effects without causing non-identifiability.  We carried out an additional numerical experiment that shows that using the tree structure as a way to parametrize the model without adding random effects does not lead to improved inference. For a more detailed discussion see Supporting Information~S8.

Like other unsupervised learning methods, LTN-LDA is unable to differentiate between different scenarios giving rise to the same sampling distributions. That is, LTN-LDA, or any other models for that matter, cannot distinguish between multiple subcommunities and a single over-dispersed one if the two give rise to the same sampling distributions. Domain knowledge is necessary to identify such possibilities; traditionally, there are two strategies to incorporate such domain knowledge. The first is through modeling assumptions, such as modelling how large the single-subcommunity dispersion is through the hyperpriors on the $\tau_k^i$. The other strategy is using a decision theoretic formulation that introduces certain loss functions to carry out post-hoc merging of the identified topics.

LTN-LDA could be extended to explicitly account for outcomes as in \cite{2007Blei}.  Moreover, we believe that the idea of incorporating cross-sample heterogeneity in MM models could be valuable beyond the context of microbiome compositions. In topic models, for example, one might expect different authors to write on the same topic using different vocabulary. LTN-LDA has the potential to be applicable to these other contexts as well, though the immediate challenge is finding an appropriate tree structure.

\subsection*{Software}

The version of the dataset used in \cite{2019_Sankaran} is available at \url{https://github.com/krisrs1128/microbiome_plvm/tree/master/data/antibiotics-study}.  
Reproducible code and data for this paper is available \url{https://github.com/PatrickLeBlanc/ReproduceLTNLDAPaper}.  R code for implementing the LTN-LDA model is available in the LTNLDA package \url{https://github.com/PatrickLeBlanc/LTNLDA}. 

\section*{Acknowledgments}
LM's research is partly supported by NIGMS grant R01-GM135440 as well as NSF grants DMS-1749789 and DMS-2013930. Part of the research was completed when PL was supported by an NSF-sponsored IBIEM fellowship at Duke University.

\bibliography{References} 
\bibliographystyle{plainnat}

\clearpage

\clearpage

\clearpage

\clearpage

\clearpage

\clearpage

\clearpage

\clearpage

\section*{Supporting Information}

\subsection*{S1: DTM-LDA}

Before developing the LTN-LDA model that we propose in this paper, we had initially attempted to introduce a ``DTM-LDA'' model, which uses the DTM to model cross-sample variability within topics. However, it turns out that the DTM-LDA is actually much more computationally demanding and in fact not computationally feasible for even moderately sized data sets, and this prompted us to seek an alternative solution.

We describe the DTM-LDA model that we had initially considered, and demonstrate how the computational difficulties arise. 
Our full DTM-LDA model is
\[
\begin{aligned}
y_{d,k}(A_l)\,|\, y_{d,k}(A), \theta_{d,k}(A), z_{dn}=k &\propto {\rm Bin}(y_{d,k}(A_l)\,|\,y_{d,k}(A),\theta_{d,k}(A)) \\
z_{dn}\,|\,\phi_d &\sim {\rm Mult}(1,\phi_d)  \\
\phi_d \,|\, \alpha &\sim {\rm Dir}(\alpha)\\
\theta_{k,d}(A)\,|\,\theta(A), \tau_k(A) &\sim {\rm Beta}(\theta_k(A)\tau_k(A),(1-\theta_k(A))\tau_k(A))\\
\theta_k(A) \,|\, \theta_0(A), \tau_0(A) &\sim {\rm Beta}(\theta_0(A)\tau_0(A),(1-\theta_0(A))\tau_0(A)) \\
\log(\tau_k(A)) &\sim {\rm Unif}(1,7)
\end{aligned}
\]
for $k\in\{1,\dots,K\}$, $d\in\{1,\dots,D\}$, $n\in\{1,\dots,N_d\}$, and $A\in\mathcal{I}$. In this model, we adopted a uniform hyperprior on the log of the per topic dispersion parameter $\tau_k$, and a Beta hyperprior for the per topic mean parameter $\theta_k$. Note that the computational issues will remain the same no matter which hyperpriors one adopt for these parameters as there are no known conjugate priors for the DT distribution.

Let $\z$ denote a vector encompassing all subcommunity assignments from all samples, $\w$ denote a vector encompassing all sequencing reads from all samples,the superscript $-(d,n)$ indicate that the $n^{th}$ read in the $d^{th}$ sample is excluded, and $\mathcal{P}_{w_{d,n}}$ be a path leading from the root node $\mathcal{R}$ of $\mathcal{T}$ to the leaf corresponding to the sequencing read $w_{d,n}$.  Then the form of the full conditional for updating the subcommunity assignments is
\[
\begin{aligned}
p(&z_{d,n}=k' \,| \,\z^{-(d,n)},\w) \propto (y_{d,k'}(\mathcal{R})^{-(d,n)}+\alpha_k) \\
&\times \frac{\prod_{A\in\mathcal{P}_{w_{d,n}}}  \int \int  \prod_{d=1}^D \bigg[\frac{B(y_{d,k'}(A_l)+1+\theta_{k'}\tau_{k'},y_{d,k'}(A_r)+1+(1-\theta_{k'})\tau_{k'})}{B(\theta_{k'}\tau_{k'},(1-\theta_{k'})\tau_{k'})}\bigg]p(\tau_{k'})p(\theta_{k'}|\tau_0,\theta_0) d\theta_{k'} d\tau_{k'}}{\prod_{A\in\mathcal{P}_{w_{d,n}}}  \int \int  \prod_{d=1}^D \bigg[\frac{B(y_{d,k'}(A_l)+\theta_{k'}\tau_{k'},y_{d,k'}(A_r)+(1-\theta_{k'})\tau_{k'})}{B(\theta_{k'}\tau_{k'},(1-\theta_{k'})\tau_{k'})}\bigg]p(\tau_{k'})p(\theta_{k'}|\tau_0,\theta_0) d\theta_{k'} d\tau_{k'}}.
\end{aligned}
\] 
There is no closed-form expression for the full conditional, and instead we numerically evaluate the double integral by quadrature.
While each individual integral can be computed quickly, for each iteration of the Gibbs sampler we must compute $2\times\mathcal{P}_{w_{d,n}}\times D \times \bar{N}_d \times K$ of them, where $\bar{N}_d$ is the average number of sequencing reads per document.  This results in a Gibbs sampler which is orders of magnitudes slower than the Gibbs sampler for LTN-LDA.  

\clearpage

\subsection*{S2: Block LTN-LDA}

We considered more complex covariance priors, but more flexible covariance structures do not lead to improved performance and can cause non-identifiability in the model.
For demonstration, we implemented the following model, termed Block LTN-LDA, which incorporates a block-diagonal covariance rather than a diagonal covariance in order to maintain identifiability while allowing a bit more flexibility, 
\[
\begin{aligned}
y_{d,k}(A_l)\,|\,y_{d,k}(A),\psi_{d,k}(A) &\ind {\rm Bin}(y_{d,k}(A),\theta_{d,k}(A)) \\
z_{d,n}\,|\,\bm{\phi}_d &\ind {\rm Mult}(1,\bm{\phi}_d)\\
\bm{\phi}_d \,|\, \bm{\alpha} &\overset{\text{iid}}{\sim} {\rm Dir}(\bm{\alpha})\\
\bm{\psi}_{d,k}\,|\,\bm{\mu}_k,\Sigma_k &\ind {\rm MVN}(\bm{\mu}_k,\Sigma_k), \\
\bm{\mu}_k \,|\, \bm{\mu}_0, \Lambda_0 &\overset{\text{iid}}{\sim} {\rm MVN}(\bm{\mu}_0,\Lambda_0) \\
\Sigma_k \,|\, {\rm G} &\overset{\text{iid}}{\sim} {\rm G}
\end{aligned}
\]
for $d = 1,\dots,D$, $k=1,\dots,K$, $n=1,\dots,N_d$, and $A\in \mathcal{I}$.  The form of the model is the same except that G now takes the form of a block covariance prior on $\Sigma_k$.  That is, let $\Sigma_k^U$ correspond to the subset of nodes in the upper part of the tree --- the set $\{A\in\mathcal{I}||A|\geq C\}$ --- and let $\Sigma_k^L$ correspond to the subset of nodes in the lower part of the tree --- the set $\{A\in\mathcal{I}||A| < C\}$.  The prior on $\Sigma_k^U$ we adopt has the form
\[
\begin{aligned}
\Sigma_k^U \,|\, \tau_k^U&= \text{diag}(\tau_{k}^U)  \\
\tau_{i,k}^U\,|\,a^U,b^U & \sim \text{IG}(a^U,b^U),
\end{aligned}
\]
as in LTN-LDA. In contrast, we model $\Sigma_k^L$ as
\[
\begin{aligned}
\W_k^L = (\Sigma_k^L)^{-1}\, | \, G_k^L &\sim \text{GWish}_{G_k^L}(a^L +p^L+2, b_L+\Phi^L),
\end{aligned}
\]
where $p^L$ is the number of nodes in $\Sigma_k^L$ and $(a^L,b^L) = (100,200)$.  We draw the precision $\Omega_k^L$ of the lower block covariance matrix from a G-Wishart distribution (\citesupp{2011LenkoskiDobra}).  The G-Wishart prior is a suitable covariance prior because it uses a graph to model the dependency structure and so can learn the conditional independence structure of the nodes from the data.  Moreover, unlike other Gaussian graphical models such as the Bayesian Graphical Lasso (\citesupp{2012Wang}), the G-Wishart prior allows us to concentrate the prior anywhere in the real line and control the degree of concentration.  This allows us to set the expected level of covariance appropriately while also restraining the posterior values from growing too flexible.  

Block LTN-LDA admits a Gibbs sampler similar to LTN-LDA in that every parameter except for $\Sigma_k$ has the same full conditional.  To sample the full conditional for $\Sigma_k^L$, we implement the trans-dimensional MCMC sampler described in \citesupp{2015MohammadiWit} and make use of the direct G-Wishart sampler described in \citesupp{2013Lenkoski}.    However, due to the added complexity in this full conditional, the Gibbs sampler for Block LTN-LDA is significantly slower than the one for LTN-LDA, taking approximately five times as long to complete on datasets of the size used in the paper.  

\begin{figure}
    \centering
    \includegraphics[width = \textwidth]{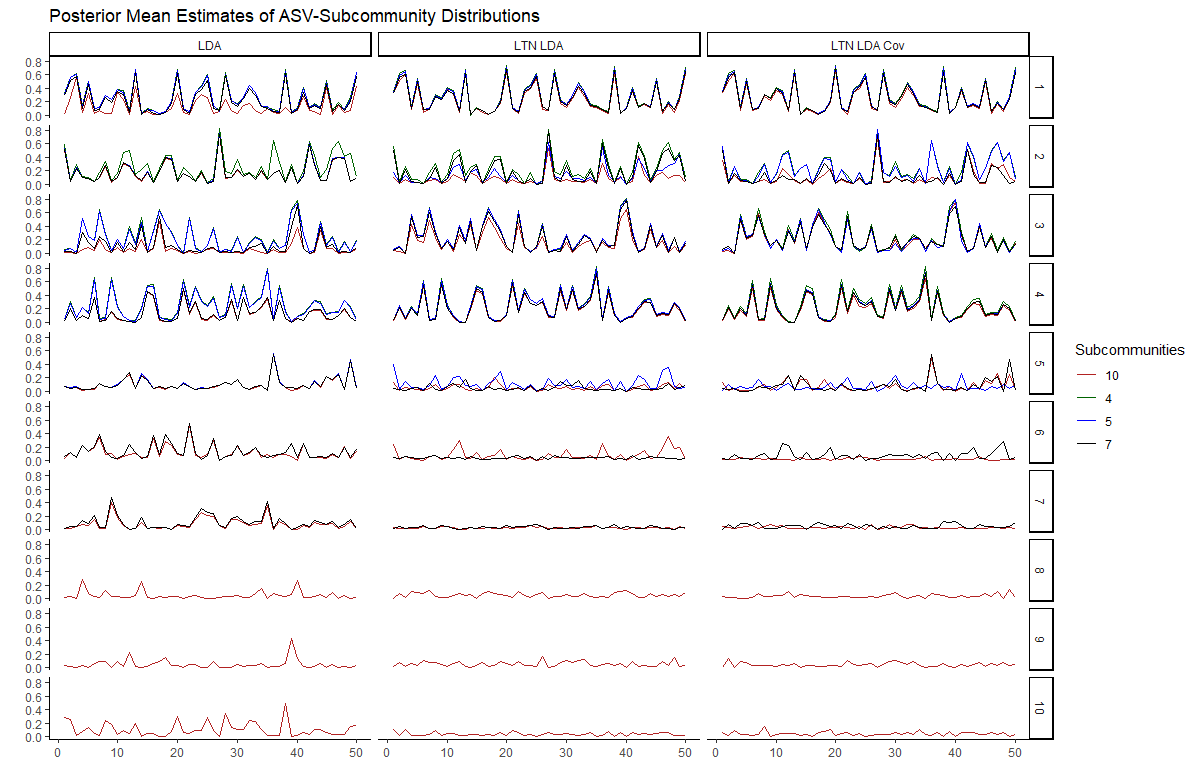}
    \caption{Posterior mean estimates for $\phi$ as $K$ varies for LDA, LTN-LDA, and Block LTN-LDA.}
    \label{fig:sim_block_cov}
\end{figure}

Despite having a more flexible covariance structure, however, Block LTN-LDA did not result in meaningfully better inference than LTN-LDA in our numerical experiments.  Specifically, we repeated the analysis in Section $3.1$ but simulated from Block LTN-LDA with a prior probability of $\frac{1}{4}$ that two nodes were dependent; all other parameters remained the same.  We then ran LDA, LTN-LDA, and Block LTN-LDA on the dataset for varying $K$ and true $C$.  The results are presented in Figure~$\ref{fig:sim_block_cov}$.  LDA behaves similarly on a Block LTN-LDA dataset as it does on an LTN-LDA dataset.  However, LTN-LDA and Block LTN-LDA predict similar mean posterior $\phi_d$ as $K$ changes despite Block LTN-LDA having generated the data.  This occurs because LTN-LDA offers a flexible enough covariance structure to capture the cross-sample heterogeneities in the data even though it assumes the nodes are independent: the additional flexibility provided by the G-Wishart priors on the lower block of the covariance matrix does not meaningfully improve the models ability to adapt itself across samples.  Given that Block LTN-LDA does not seem to provide meaningfullly more accurate inference than LTN-LDA even when it is true, and that is substantially less computationally efficient, we deem that the more flexible covariance structure does not lead to improved performance.  

\bibliographystylesupp{plainnat}
\bibliographysupp{References}

\clearpage

\subsection*{S3: Robustness to misspecified trees}

\begin{figure}
    \centering
    \includegraphics[width = \textwidth]{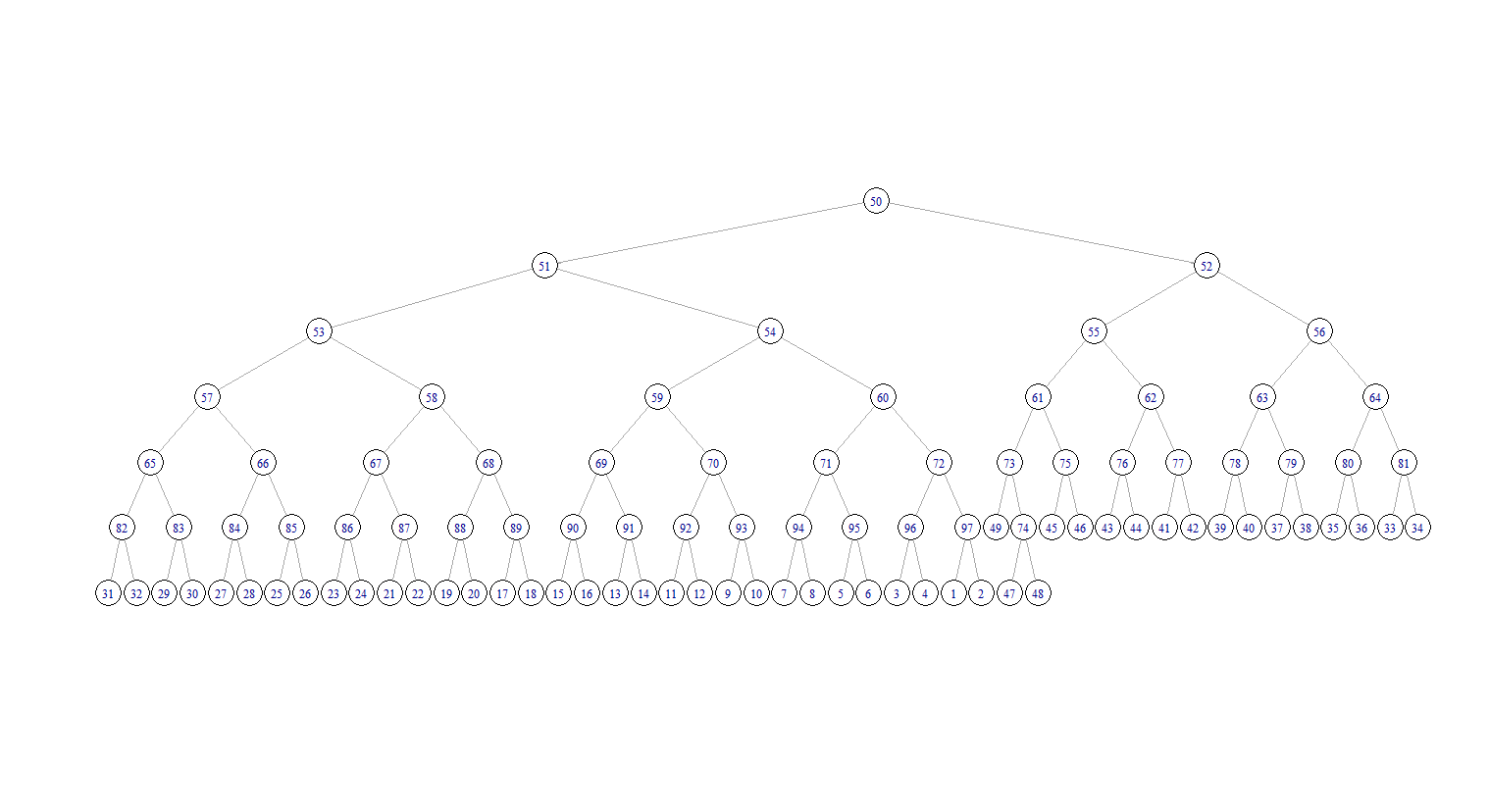}
    \caption{A close to uniform tree constructed from nodes $1,2,\dots,49$.}
    \label{fig:sim_tree_unif}
\end{figure}

\begin{figure}
    \centering
    \includegraphics[width = \textwidth]{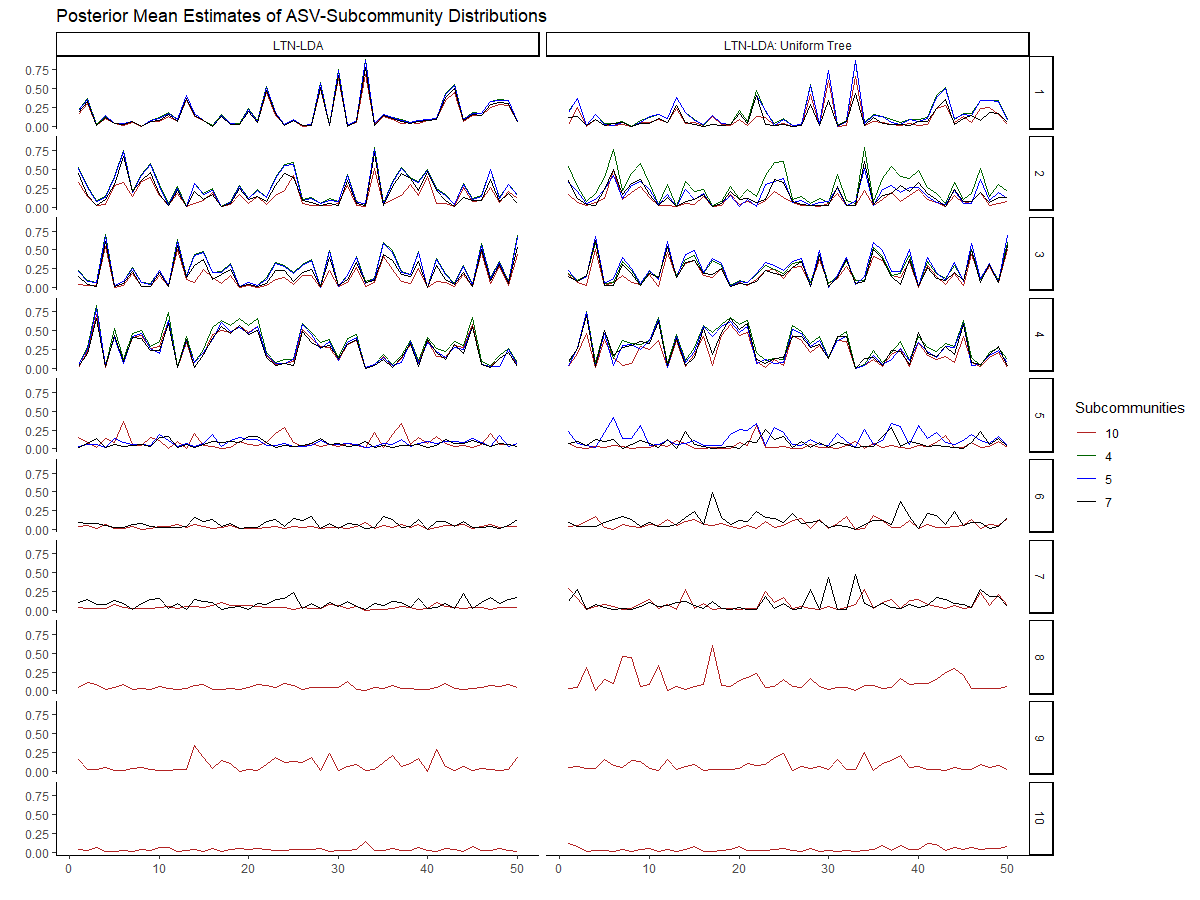}
    \caption{Posterior mean estimates for $\phi$ as $K$ varies for LTN-LDA using a ``true" tree and a uniform tree.}
    \label{fig:sim_tree_unif_phi}
\end{figure}

\begin{figure}
    \centering
    \includegraphics[width = \textwidth]{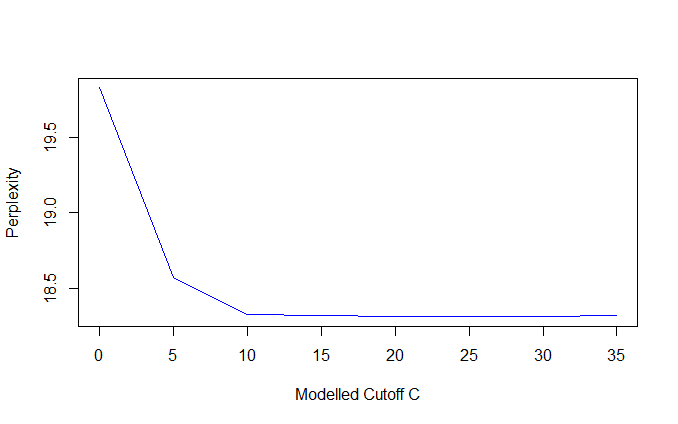}
    \caption{Perplexity results for the mispecified tree as $C$ varies}
    \label{fig:sim_trees_perp}
\end{figure}

\begin{figure}[!h]
    \centering
    \includegraphics[width = \textwidth]{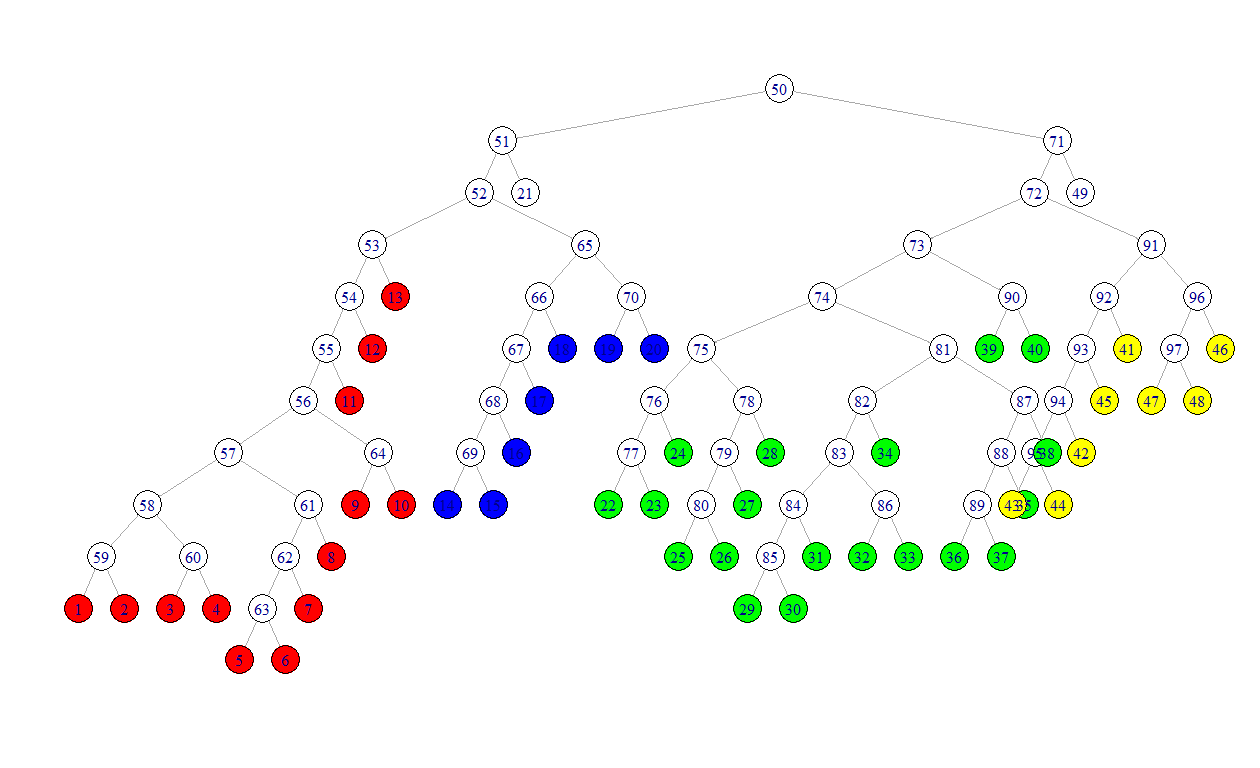}
    \caption{The phylogenetic tree used in simulations}
    \label{fig:sim_tree}
\end{figure}

The tree structure is vital to the way in which LTN-LDA models cross-sample heterogeneity, and thus it is important to investigate the robustness of the inference to the choice of the tree. To demonstrate this, we generate a dataset as in section 3.1 based on the tree given in Figure~$\ref{fig:sim_tree}$, and then repeated the analysis comparing LTN-LDA based on this correct tree to LTN-LDA using a misspecified tree as given in Figure~$\ref{fig:sim_tree_unif}$. The results are presented in Figure~$\ref{fig:sim_tree_unif_phi}$.  The inference provided by the two approaches is similar when $K = 4$, the true value.  However, as $K$ increases the mispecified tree's inference deteriorates faster than does the true tree's.  Further, we generated a dataset using LTN-LDA with the tree in Figure~\ref{fig:sim_tree} and ran a perplexity analysis as $C$ varies with the tree in Figure~\ref{fig:sim_tree_unif} as the tree.  The results are in the tree in Figure~\ref{fig:sim_trees_perp}.  We can see that the bend in the curve appears to occur before the true value, and so using the mispecified tree can also influence the choice of the tuning parameter. On the other hand, the fitted subcommunity abundances generally maintains the same shape for misspecified $K$ and $C$, indicating a level of robustness of LTN-LDA with respect to the choice of the tree.

\clearpage

\subsection*{S4: Collapsed blocked Gibbs sampler}

We integrate the $\bm{\phi}_d$ out of the sampling model and proceed with a collapsed Gibbs sampler to improve convergence  \citep{2004_Griffiths}. The full conditionals we will sample from are thus 
\[
\begin{aligned}
(1) & \qquad (\bm{v}_{d,k},\z_d) \,|\, \bm{\psi}_{d,k},\bm{\mu}_k,\Sigma_k, \Lambda_0, \bm{\alpha}, a_1, a_2, b \ind p(\bm{v_{d,k}},\z_d|\bm{\psi}_{d,k}, \bm{\alpha}) \\
(2) & \qquad \psi_{d,k} \,|\, \bm{v}_{d,k},\z_d,\bm{\mu}_k,\Sigma_k,\Lambda_0, \bm{\alpha}, a_1, a_2, b \ind p(\bm{\psi}_{d,k}|\bm{v}_{d,k},\z_d,\bm{\mu}_k,\Sigma_k) \\
(3) & \qquad \bm{\mu}_k\,|\,\bm{v}_{d,k},\z_d,\bm{\psi}_{d,k},\Sigma_k, \Lambda_0, \bm{\alpha}, a_1, a_2, b  \ind p(\bm{\mu}_k|\bm{\psi}_{d,k},\Sigma_k, \Lambda_0) \\
(4) & \qquad \Sigma_k\,|\,\bm{v}_{d,k},\z_d,\bm{\psi},\bm{\mu}_k, \Lambda_0, \bm{\alpha}, a_1, a_2, b \ind p(\Sigma_k|\bm{\psi},\bm{\mu}_k, \Lambda_0, a_1, a_2, b), \\
\end{aligned}
\]
The joint full conditional $(\bm{v_{d,k}},\z_d)$ is
\[
\begin{aligned}
\z_d \,|\, \bm{\psi}_{d,k}, \bm{\alpha}  &\ind p(\z_d|\bm{\psi}_{d,k}, \bm{\alpha} ) \\
\bm{v_{d,k}} \,|\, \z_d, \bm{\psi}_{d,k}, \bm{\alpha} &\ind p(\bm{v_{d,k}}|\z_d, \bm{\psi}_{d,k} ).
\end{aligned}
\]
To sample the vector $\z_d$ from its full conditional, we sample each subcommunity assignment in order from its multinomial full conditional: 
\[
p(z_{d,n} = k|\z_{d}^{-n},\bm{\psi}_{d,k},\bm{\alpha}) \propto (y_{d,k}(\mathcal{R})^{-n} + \alpha)\times \beta_{k}^{w_{d,n}}, 
\]
where $\z_{d}^{-n}$ is the vector of all subcommunity assignments in sample $d$ except for $z_{d,n}$ and $y_{d,k}(\mathcal{R})^{-n}$ is the number of sequencing reads in sample $d$ descended from the root node $\mathcal{R}$ assigned to subcommunity $k$ not counting the $n^{th}$ token.  To sample from the full conditional for $\bm{v_{d,k}}$, we draw  $v_{d,k}(A)$ for each $A\in\mathcal{I}$:
\[
v_{d,k}(A)\,|\,y_{d,k}(A),\psi_{d,k}(A) \ind {\rm PG}(y_{d,k}(A),\psi_{d,k}(A)),
\]
the conjugate full conditional of a P\'olya-Gamma distribution derived in \cite{2013_Polson}.  However, existing P\'olya-Gamma samplers are slow for the current context and so for $y_{d,k}(A) \geq 30$ we use an approximate P\'olya-Gamma sampler proposed in \cite{2019_Glynn}, which uses the Central Limit Theorem to approximate a normal distribution:
\[
{\rm N}\bigg(
\frac{y_{d,k}(A)^2}{2\psi_{d,k}(A)} \text{tanh}\bigg(\frac{\psi_{d,k}(A)}{2}\bigg),
\frac{y_{d,k}(A)^2}{4\psi_{d,k}(A)^3} \text{sech}^2\bigg(\frac{\psi_{d,k}(A)}{2}\bigg) (\text{sinh}(\psi_{d,k}(A)) - \psi_{d,k}(A))
\bigg).
\]
The full conditionals for $\bm{\mu}_k$ and the $\tau_{i,k}$ follow by conjugate updating:
\[
\begin{aligned}
\bm{\mu}_k \,|\,\bm{\psi}_{d,k},\Sigma_k, \Lambda_0 &\ind {\rm MVN}\left((\Lambda_0^{-1}+D\Sigma_k^{-1})^{-1}\Sigma_k^{-1}\sum_{d=1}^D \psi_{d,k},(\Lambda_0^{-1}+D\Sigma_k^{-1})^{-1}\right) \\
\tau_{i,k}\, |\,\bm{\psi},\bm{\mu}_k,, a_1, a_2, b &\ind {\rm IG}\left(a_1 + \frac{D}{2}, \frac{2b + \sum_{d=1}^D(\psi_{d,k}(A_i)-\mu_k(A_i)^2)}{2}\right)\ \text{ if } |A_i| \geq C \\
\tau_{i,k} \,|\, \bm{\psi},\bm{\mu}_k,, a_1, a_2, b &\ind {\rm IG}\left(a_2 + \frac{D}{2}, \frac{2b + \sum_{d=1}^D(\psi_{d,k}(A_i)-\mu_k(A_i)^2)}{2}\right)\ \text{ if } |A_i| < C.
\end{aligned}
\]
Further, the full conditional for $\bm{\psi}_{d,k}$ is also normal,
\[
\bm{\psi}_{d,k}\,|\, \z_d, \bm{v}_{d,k},\mu_k, \Sigma_k \ind {\rm MVN}\left( (\Sigma_k^{-1}+\text{diag}(\bm{v}_{d,k}))^{-1}(\Sigma_k^{-1}\bm{\mu}_k + \bm{\kappa}_{d,k}),(\Sigma_k^{-1}+\text{diag}(\bm{v}_{d,k}))^{-1}\right).
\]

\begin{figure}
\begin{subfigure}{0.24\textwidth}
    \centering
    \includegraphics[width = 0.95\linewidth]{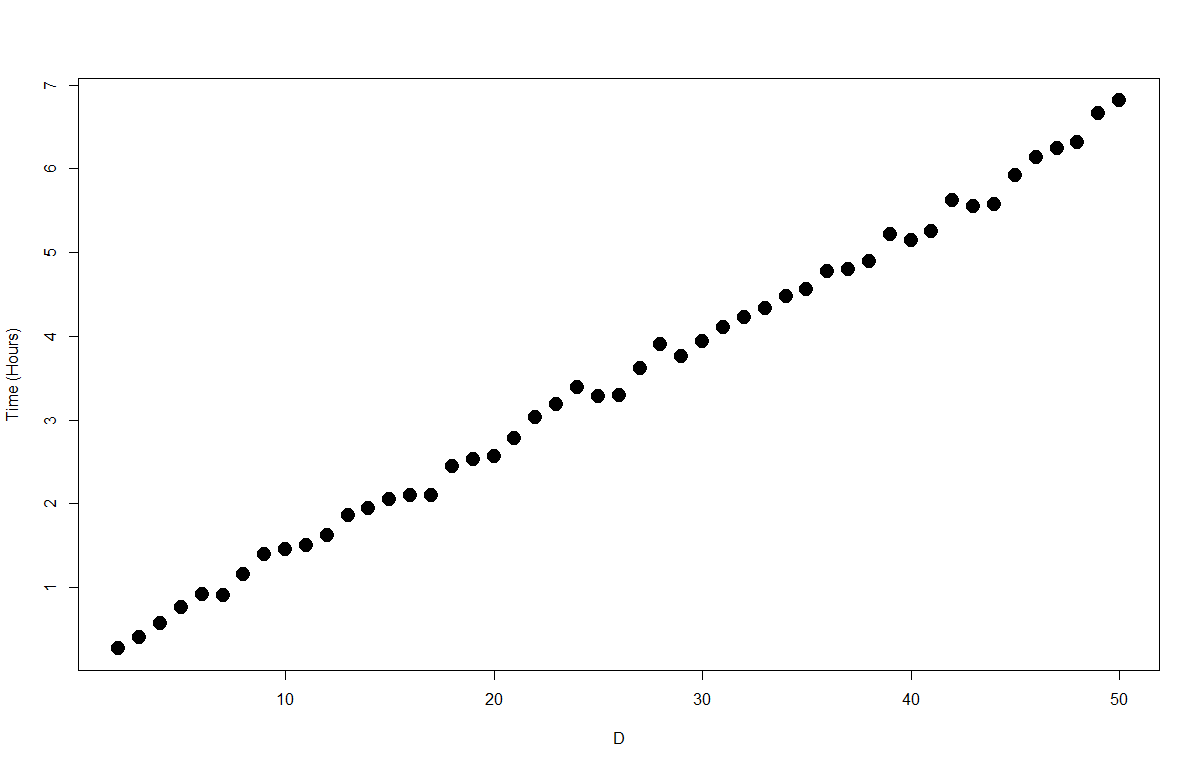}
    \caption{Scaling with D}
    \label{fig:sim_time_d}
\end{subfigure}
\begin{subfigure}{0.24\textwidth}
    \centering
    \includegraphics[width = 0.95\linewidth]{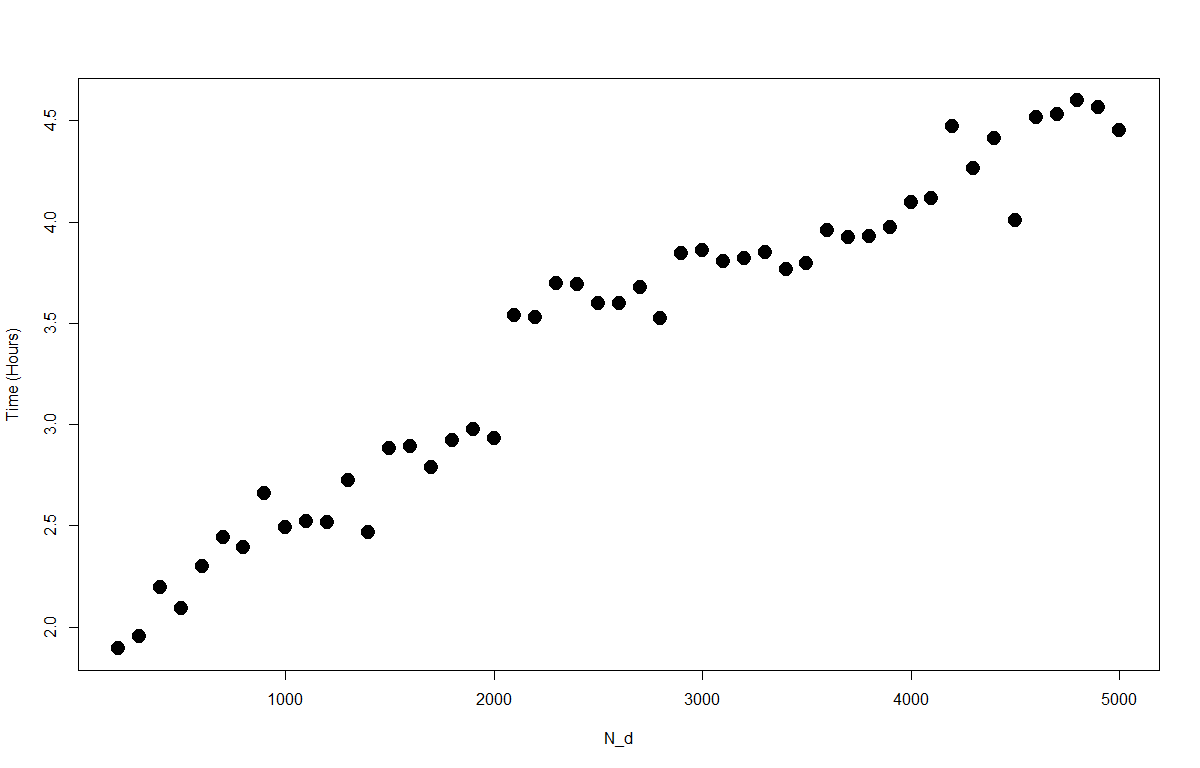}
    \caption{Scaling with $N_d$}
    \label{fig:sim_time_n}
\end{subfigure}
\begin{subfigure}{0.24\textwidth}
    \centering
    \includegraphics[width = 0.95\linewidth]{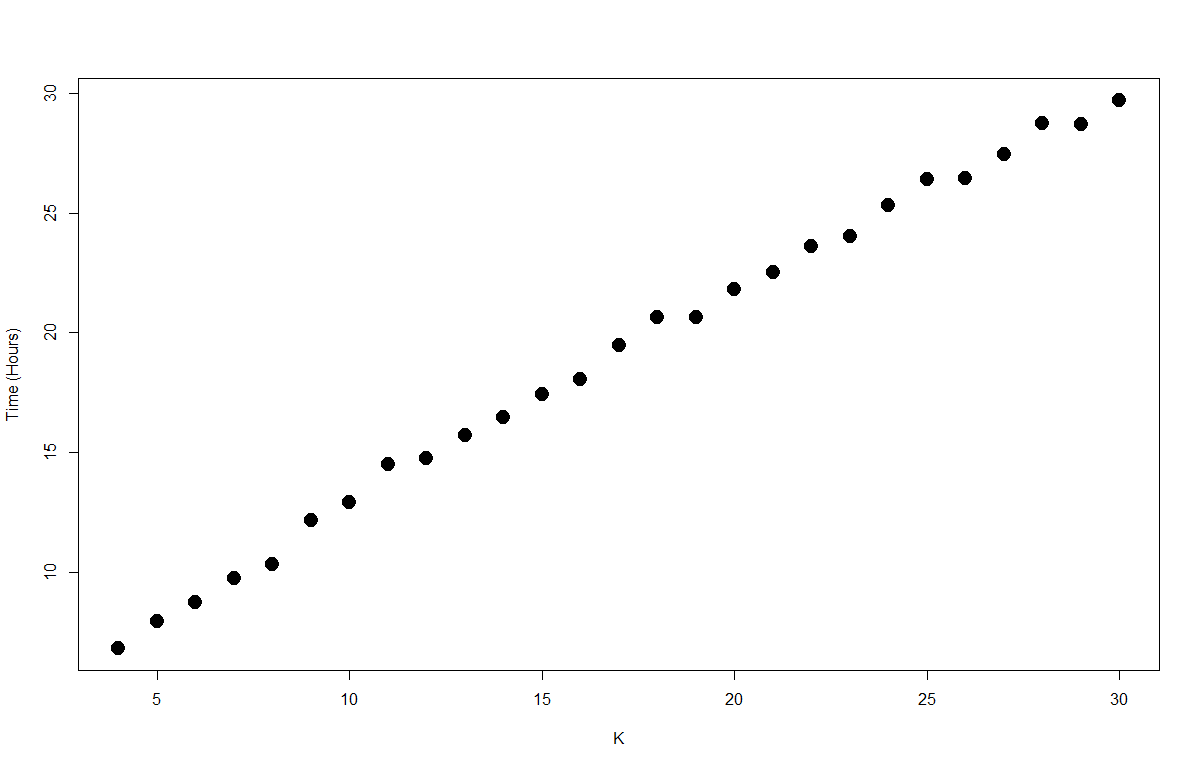}
    \caption{Scaling with K}
    \label{fig:sim_time_k}
\end{subfigure}
\begin{subfigure}{0.24\textwidth}
    \centering
    \includegraphics[width = 0.95\linewidth]{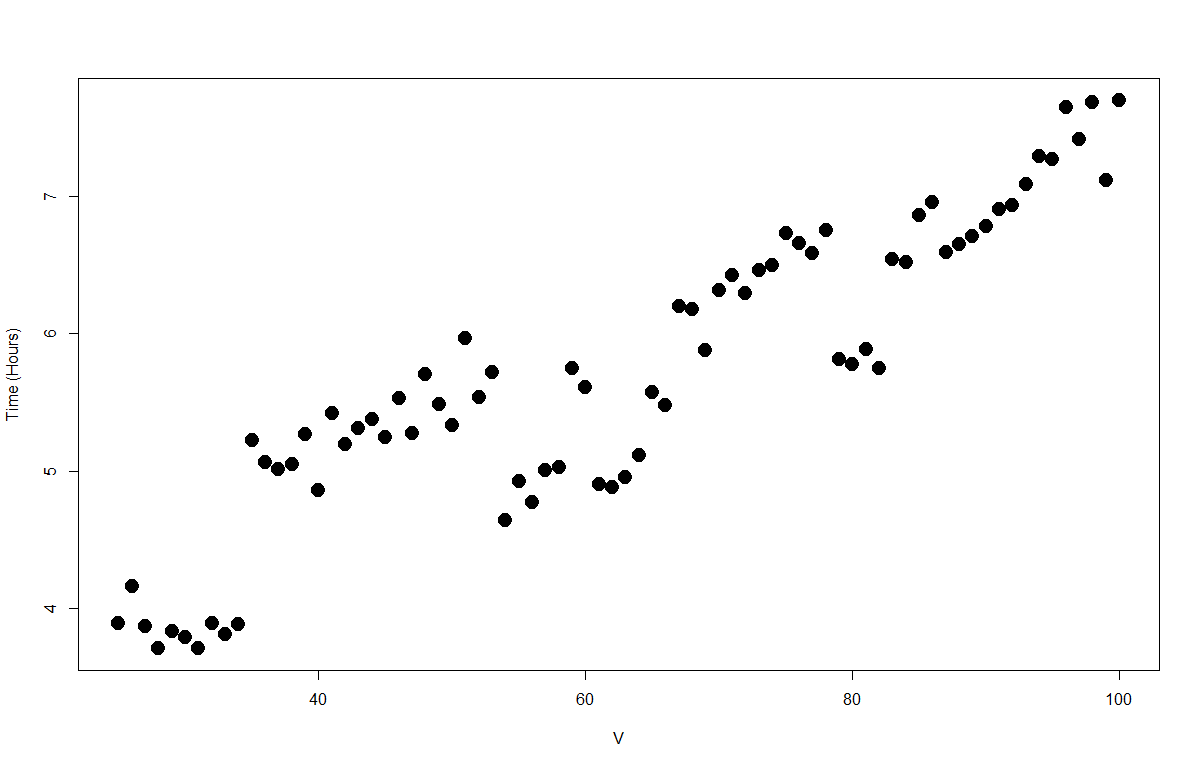}    \caption{Scaling with V}
    \label{fig:sim_time_v}
\end{subfigure}
\caption{Scalings}
\label{fig:scaling}
\end{figure}

The Gibbs sampling algorithm scales linearly with $D$ (Figure~$\ref{fig:sim_time_d}$), $N_d$ (Figure~$\ref{fig:sim_time_n}$), $K$ (Figure~$\ref{fig:sim_time_k}$), and $V$ (Figure~$\ref{fig:sim_time_v}$). The computation time does not scale with the tree parameters $\mathcal{T}$ and $C$ because of the diagonal covariance structure.

\clearpage

\subsection*{S5: The phylogenetic tree used in the simulation study}

\clearpage

\subsection*{S6: Perplexity}

Perplexity is a transformation of predictive log-likelihood commonly used to assess topic models.  If $p(\w_{d^{(te)}}|\mathcal{M})$ is the predictive log-likelihood of a test set sample $d^{(te)}$ given a collection of parameters $\mathcal{M}$, then perplexity is defined as
\[
\exp\bigg(-\frac{\sum_d p(\w_{d^{(te)}}|\mathcal{M})}{\sum_d N_{d^(te)}} \bigg),
\]
where $N_{d^(te)}$ is the number of sequencing reads in $d^{(te)}$

The document completion method for computing perplexity for LDA is described in section $5$ of \cite{2009_Wallach}.  It involves splitting each sample $d^{(te)}$ in the test set into two halves, $d^{(te),1}$ and $d^{(te),2}$.  A modified Gibbs sampler is run on the first half, $d^{(te),1}$ with the value of $\beta_k$ set equal to the posterior mean of $\beta_k$ on the training set.  The results of this Gibbs sampler are used to develop estimates for $\phi_d$ and then for perplexity.

We modify this procedure for LTN-LDA.  A modified Gibbs is run on $d^{(te),1}$, fixing the values of $\mu_k$ and $\Sigma_k$ at their posterior means from the training set.  If there are $I$ iterations in the Gibbs sampler, then the estimate of $\bm{\phi}_d$ at iterate $i$ is $\bm{\phi}_d(i)$ and the estimate of $\bm{\beta}_{d,k}$ of iterate $i$ is $\bm{\beta}_{d,k}(i)$.  We can then take a Monte Carlo estimate over all ASVs observed in $d^{(te),2}$ to estimate the predictive likelihood of $d^{te}$:
\[
\frac{1}{I}\sum_{i=1}^I \sum_{n \in \w_{d^{(te),2}}} \log\bigg( \sum_{k=1}^K \phi_{d}^{k}(i)\beta_{d,k}^{w_{d,n}}(i)\bigg).
\]
This procedure can be repeated for every sample in the test set, and the resulting set of predictive likelihood estimates can be transformed into a perplexity estimate.  

\clearpage

\subsection*{S7: Dethlefsen and Relman Tree}

\begin{figure}[h]
    \centering
    \includegraphics[width = \textwidth]{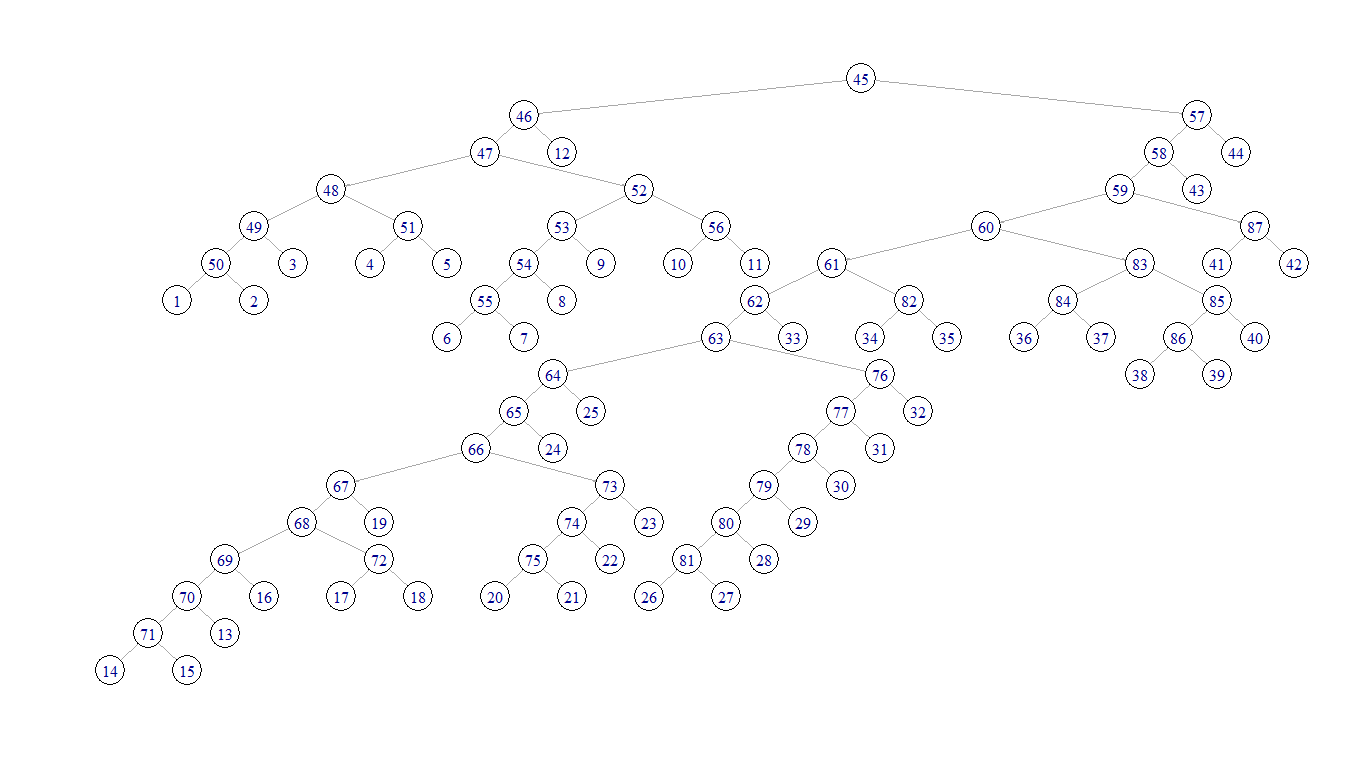}
    \caption{The tree resulting from the dataset of \cite{2011_Dethlefsen}}
    \label{fig:sank_tree}
\end{figure}

\subsection*{S8: Separating the effects of the tree from that of the random effect}

LTN-LDA incorporates two new effects: the tree structue and random effects in cross-sample heterogeneity. We note that using the tree structure alone without allowing random effects does not improve inference in any way. On the other hand, the tree structures provides guidance on how to parsimoniously model the random effects without causing non-identifiability. Moreoever, without the tree structure, and hence without computational techniques such as the PG augmentation, there is no known scalable sampling algorithm that allows us the compute the resulting posterior model efficiently.

\begin{figure}
     \centering
     \includegraphics[width = \textwidth]{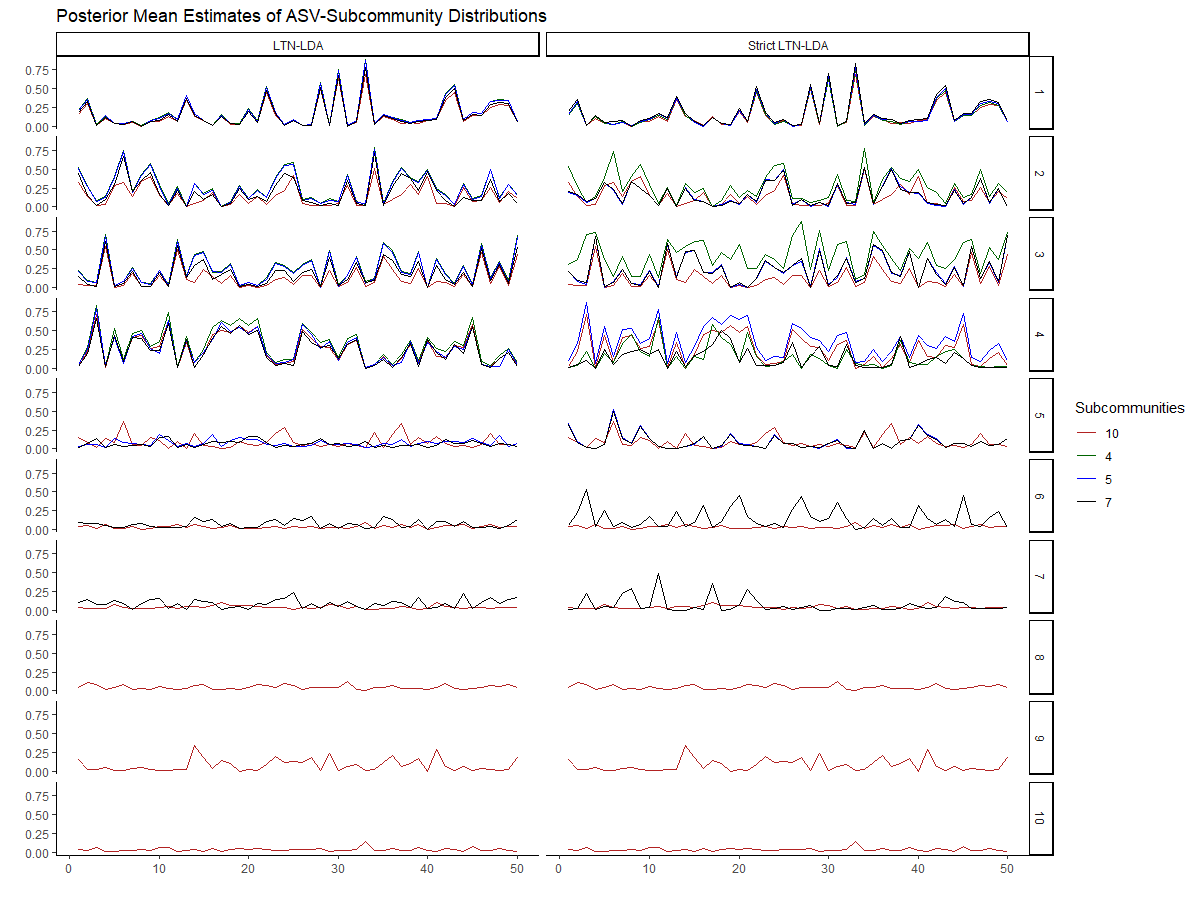}
     \caption{Posterior mean estimates for $\phi$ as $K$ varies for LTN-LDA with recommended covariance priors vs those with ``knock-out'' covariance priors.}
     \label{fig:sim_strict_priors}
 \end{figure}

We provide evidence that the tree structure alone without the random effects does not improve the inference over LDA.  We ``knock out'' the random effects by forcing the sample-specific distributions $\beta_{d,k}$ to not vary from sample-to-sample.  Thus, we can approximate this model with an existing Gibbs sampler. We then replicate the results of Section $3.1$ of the manuscript but comparing LTN-LDA with the usual prior to LTN-LDA with this ``knock-out'' covariance prior, and present the results in Figure~$\ref{fig:sim_strict_priors}$.  The version with strict priors misestimates subcommunity proportions for $K = 4$, and splits the subcommunities as $K$ grows.  We thus deduce that using the tree structure without allowing cross-sample heterogeneity does not reproduce the positive results in the paper.  

However, it is difficult to implement a model with just random effects without a tree structures. On the modeling side, without the tree structure, one must induce more complex constraints on the covariance to ensure identifiability. Moreover, without the tree structure, it is unclear how these models can be implemented efficiently to be applicable to modern microbiome data sets.  It would be interesting to see such a comparison so that we can understand to what extent to improvement is due to the tree modeling assumption. However, we know of no such existing implementation of the models suggested. Note that even the seemingly simple Dirichlet random effect model will require a high-dimensional ($m$-dim) numerical integral where $m$ is the number of taxa within each iteration for the same reason that the DTM requires numerical integration as we mentioned before---there is no known conjugate priors for the parameters in the Dirichlet distribution. For these reasons, we feel that such a comparison goes beyond the scope of this manuscript. Finally, we emphasize that it is indeed the adoption of a tree structure that provides an efficient means to computing. We believe in this regard our use of the tree goes beyond  prior works that uses the tree only for modeling purposes, not a computational technique.
\clearpage

\end{document}